\documentclass[fleqn,usenatbib]{mnras}
\usepackage{amssymb}
\usepackage{graphicx}
\usepackage{amsmath}
\usepackage[T1]{fontenc}
\usepackage{ae,aecompl}
\usepackage{newtxtext,newtxmath}
\usepackage{multirow}

\title[The UV LF at $z\simeq11$]
{The galaxy UV luminosity function at $\mathbf{z \simeq 11}$ from a suite of public JWST ERS, ERO and Cycle-1 programs}
\author[D. J. ~McLeod et al.]
{D. J. McLeod$^{1}$\thanks{Email: mcleod@roe.ac.uk},  C. T. Donnan$^{1}$, R. J. McLure$^{1}$,
  J. S. Dunlop$^{1}$, D. Magee$^{2}$, R. Begley$^{1}$, \and A. C. Carnall$^{1}$, F. Cullen$^{1}$, R. S. Ellis$^{3}$, M. L. Hamadouche$^{1}$, T. M. Stanton$^{1}$
\footnotesize\\\\
$^{1}$ Institute for Astronomy, University of Edinburgh, Royal Observatory, Edinburgh EH9 3HJ, UK\\
$^{2}$ Department of Astronomy and Astrophysics, UCO/Lick Observatory, University of California, Santa Cruz, CA 95064, USA\\
$^{3}$ Department of Physics and Astronomy, University College London, Gower Street, London, WC1E 6BT, UK
}

\date{Accepted XXX. Received YYY; in original form ZZZ}

\pubyear{2023}

\begin{document}
\label{firstpage}
\pagerange{\pageref{firstpage}--\pageref{lastpage}}
\maketitle

\begin{abstract}
We present a new determination of the evolving galaxy UV luminosity function (LF) over the redshift range $9.5<z<12.5$ based on a wide-area ($>250$\,arcmin$^2$) data set of {\it JWST} NIRCam near-infrared imaging assembled from thirteen public \textit{JWST} surveys. Our relatively large-area search allows us to uncover a sample of 61 robust $z>9.5$ candidates detected at $\geq 8\sigma$, and hence place new constraints on the intermediate-to-bright end of the UV LF. When combined with our previous \textit{JWST}+UltraVISTA results, this allows us to measure the form of the LF over a luminosity range corresponding to four magnitudes ($M_{1500}$). At these early times we find that the galaxy UV LF is best described by a double power-law function, consistent with results obtained from recent ground-based and early \textit{JWST} studies at similar redshifts. Our measurements provide further evidence for a relative lack of evolution at the bright-end of the UV LF at $z=9-11$, but do favour a steep faint-end slope ($\alpha\leq-2$). The luminosity-weighted integral of our evolving UV LF provides further evidence for a gradual, smooth (exponential) decline in co-moving star-formation rate density ($\rho_{\mathrm{SFR}}$) at least out to $z\simeq12$, with our determination of $\rho_{\mathrm{SFR}}(z=11)$ lying significantly above the predictions of many theoretical models of galaxy evolution.
\end{abstract}

\section{INTRODUCTION}
Although \textit{JWST} has not yet completed its first year of science operations, it is already transforming our view of the young Universe. Despite its many achievements, the \textit{Hubble Space Telescope} (\textit{HST}) was unable to advance the search for early galaxies significantly beyond redshifts $z \simeq 10$, due to its limited near-infrared wavelength coverage ($\lambda < 1.6$\,$\mu$m)  (e.g., \citealt{Ellis2013,Coe2013,Oesch2016}). By contrast, the exquisite near/mid-infrared imaging now being provided by the NIRCam instrument on-board the larger and colder {\it JWST} has already pushed the redshift frontier out to $z \simeq 13$ \citep{Naidu2022,Finkelstein2022, Robertson2023}, with candidate high-redshift galaxies already being uncovered at redshifts as extreme as $z=16-17$ \citep{Harikane2023}.

The photometric selection of high-redshift galaxy candidates from deep imaging not only supplies targets for spectroscopic follow-up to determine their astrophysical properties, but also enables the statistical study of the evolution of the galaxy population provided sufficiently large and robust samples can be assembled. In particular, a key goal is to chart the evolution of the galaxy luminosity function (LF) back to early times for comparison with theoretical predictions. At very high redshifts, near/mid-infrared imaging enables the evolving rest-frame {\it ultraviolet} (UV) galaxy LF to be determined, the luminosity-weighted integral of which gives the evolution of (co-moving) UV luminosity density, which can then be converted into the evolution of cosmic star-formation rate density, $\rho_{\mathrm{SFR}}$. In large part due to the impact of {\it HST}, there is now a general consensus over the evolution of  $\rho_{\mathrm{SFR}}$ out to $z \simeq 8$. However, before the advent of {\it JWST}, the lack of significant galaxy samples at higher redshifts resulted in considerable disagreement over the evolution of star-formation rate density at earlier times. Specifically, some studies concluded in favour of a rapidly steepening decline in $\rho_{\mathrm{SFR}}$ with increasing look-back time beyond $z \simeq 8$ (e.g., \citealt{Oesch2013,Oesch2018}), while others presented evidence for a continued smooth, more gradual evolution implying the existence of substantial star-formation activity at still earlier times \citep{McLeod2015,McLeod2016}. 

This debate/disagreement over the very high-redshift evolution of the UV LF, and hence $\rho_{\mathrm{SFR}}$, was largely a consequence of the limitations of pre-\textit{JWST} facilities mentioned above. Specifically, at redshifts $z \geq 9$ the WFC3/IR camera on {\it HST} rapidly runs out of filters capable of sampling continuum emission beyond the Lyman break at $\lambda_{\mathrm{rest}}=1216$\,\AA. For most of the deep extragalactic imaging surveys performed with {\it HST} only F125W and F160W imaging are available for this purpose, at least for the majority of the CANDELS \citep{Grogin2011} area. The addition of the F140W filter for such surveys as the HUDF \citep{Beckwith2006, Ellis2013}, CLASH \citep{Postman2012} and the Frontier Fields \citep{Lotz2017} was key for selecting $z=9$ galaxies in \citet{McLeod2015,McLeod2016} as it provided a third filter with high-resolution imaging that probes beyond the Lyman break. The \textit{Spitzer} space telescope (e.g., S-CANDELS; \citealt{Ashby2015}) and the deepest ground-based $K_{s}-$band imaging (e.g. \textit{Very Large Telescope (VLT)}'s HUGS; \citealt{Fontana2014}) helped to extend the wavelength baseline beyond $\lambda = 1.6$\,$\mu$m. Using the additional wavelength coverage of \textit{Spitzer}, studies such as \cite{RobertsBorsani2016} and \citet{FinkelsteinCANDELS} provided constraints on the rest-frame optical continuum of some brighter high-redshift galaxies. The increased constraints on the rest-frame optical from \textit{Spitzer} also allows more accurate stellar mass estimates at high-redshift, as demonstrated by \cite{Stefanon2021b} who were able to constrain the galaxy stellar mass function at $z=6-10$ by leveraging \textit{HST}-selected samples with exceptionally deep IRAC data from the GREATS program \citep{Stefanon2021}. While \textit{Spitzer} had been incredibly useful for expanding our understanding of high-redshift galaxy properties, particularly for bright galaxies, this redder non-{\it HST} imaging lacked the resolution and the depth to benefit searches for fainter galaxies at $z > 8$.

\textit{JWST} has effectively unlocked the longer wavelength near/mid-infrared coverage previously only provided by ground-based facilities and \textit{Spitzer}, but now with enormous improvements in both angular resolution and depth. Several early, public \textit{JWST} NIRCam surveys have reached $5\sigma$ limiting depths of $28.5-29.0$\,mag in filters spanning the wavelength range $\lambda \simeq 1 - 5$\,$\mu$m, already probing several magnitudes deeper than typical \textit{Spitzer} surveys, and with diffraction-limited imaging down to $\lambda \simeq 1$\,$\mu$m delivering near-infrared imaging of much improved resolution compared to \textit{HST's} WFC3/IR. This combination of depth and resolution has already been transformative in the selection of ultra-high redshift galaxies. Encouragingly, many of the $z>10$ candidates photometrically selected from the early NIRCam imaging are already being confirmed spectroscopically with {\it JWST} NIRSpec \citep{ArrabalHaro2023, ArrabalHaro2023b, Harikane2023b}.

While several theoretical models of galaxy evolution appear to predict a rapid decline in $\rho_{\mathrm{SFR}}$ at $z>8$ (e.g., \citealt{Mason2015, Yung2019}), early \textit{JWST} studies have provided evidence to the contrary. \citet{Donnan2023} conducted a wide-area, ground-based  search for bright $z=8-10$ galaxy candidates using COSMOS/UltraVISTA DR5 \citep{McCracken2012}, in combination with a search for fainter galaxies at these redshifts with the \textit{JWST} CEERS \citep{Finkelstein2023}, GLASS-parallel \citep{Treu2022} and SMACS0723-73 \citep{Pontopiddan2022} public surveys. The resulting UV LF determination was found to be consistent with the earlier \textit{HST}-based works of \citet{McLeod2015, McLeod2016}, with $\rho_{\mathrm{SFR}}$ displaying an extended smooth decline at $z>8$ and indeed continuing to display a log-linear relation with redshift until at least $z\sim12$. At the bright-end, the number densities of $z=8-10$ galaxy candidates found using the latest COSMOS/UltraVISTA ground-based data suggested that the UV LF follows a double power-law functional form, as previously found by \citet{Bowler2020}. Although the exceptionally bright $z=16$ candidate CEERS-93316 first reported in \citet{Donnan2023} has now been spectroscopically confirmed to be a lower redshift interloper at $z = 4.9$ \citep{ArrabalHaro2023}, this is a particularly pathological case in which a peculiar combination of very strong emission lines conspired to masquerade as the blue continuum above the Lyman-break. Another potential $z=16$ candidate behind Stephan's Quintet may yet suggest that significant star-formation activity is still taking place at $z\geq14$ \citep{Harikane2023}.

As well as the discovery of extremely high-redshift galaxies at $z>12$ \citep{Naidu2022,Atek2023,Donnan2023,Harikane2023}, \textit{JWST} has been yielding surprisingly large numbers of bright $z\simeq10$ galaxies, with a particularly high number density reported by \citet{Castellano2023} in the area around the Abell 2744 cluster. Such densities are potentially up to a factor ten times greater than expectations based on previous observations and theoretical models of galaxy evolution. This would challenge our existing understanding, with many existing theoretical models of galaxy evolution having to invoke more efficient star formation or less dust attenuation at the highest redshifts to match these observational results. Alternatively, these early observations may be biased to a particular population of young and rapidly star-forming galaxies \citep{Mason2023}. There is thus an open question as to how accurately the early {\it JWST}-selected galaxy samples represent the true number densities of bright ${z=10}$ galaxies, with existing high-redshift galaxy searches typically restricted to small areas.

To address this issue, the study presented here aims to exploit all relevant available {\it JWST} NIRCam Early Release Observations (ERO), Early Release Science (ERS) and public Cycle-1 programs to conduct a systematic wide-area search for extreme-redshift galaxy candidates with high signal-to-noise ($\geq8\sigma$). The resulting combined survey, covering a raw area of $\simeq260$ arcmin$^2$, provides a new opportunity to uncover statistically significant samples of galaxies in the redshift range $9.5<z<12.5$, and hence yield improved constraints on the number densities of bright to intermediate $M_{1500}$ galaxies over this redshift range. By combining the new results with our previously-derived constraints on the number densities of fainter galaxies \citep{Donnan2023}, we are hence able to determine the evolving UV LF and $\rho_{\mathrm{SFR}}$ at $z=10-12$, deep into the redshift range where the results derived using \textit{HST} had diverged. Moreover, our wide-area search may also uncover other examples of extremely high-redshift ($z>12$) galaxy candidates, such as those found in the ERO surveys (e.g. \citealt{Atek2023, Harikane2023}).

The structure of this paper is as follows. In Section 2 we describe each of the data sets included in our high-redshift galaxy search. Section 3 describes the construction of our multi-wavelength catalogues and candidate selection. We discuss our candidates and draw comparisons with other recent \textit{JWST} searches in the literature in Section 4. In Section 5, we use our high-redshift galaxy sample to determine the form of the UV LF over the redshift range $9.5<z<12.5$ and how it evolves from $z \simeq 8$ to $z \simeq 12$. In Section \ref{sfrd_section} we integrate the UV LF in order to determine the star-formation rate density and how it evolves with cosmic time. Finally, in Section 7 we summarise our conclusions. Throughout the paper, we adopt a cosmology with $\Omega_{0}=0.3, \Omega_{\Lambda}=0.7$ and $H_{0}=70$ kms$^{-1}$Mpc$^{-1}$, and use the AB magnitude system  \citep{Oke1974, Oke1983}. 

\section{DATA}
\subsection{Imaging data}
\label{surveys_section}
We begin with a description of the various data sets employed in this study. Each of the data sets were downloaded as ``rate'' products, either from the Canadian Astronomy Data Centre (CADC) or STScI MAST database, and processed through the Primer Enhanced NIRCam Image Processing Library (\textsc{PENCIL}; Magee in prep) software, which is a customised version of the {\it JWST} pipeline version 1.6.2. This version of the pipeline includes additional routines for the removal of snowballs and wisps, as well as background subtraction and correcting for 1/f noise striping. Given the different timeframes in which the various data sets were taken, released, reduced and analysed, there are very slight differences in the CRDS context depending on when the data set was released, although the pmap is always at least pmap 0989, which included the most recent zero-point corrections at the time of writing. Where applicable, \textit{HST} data sets were downloaded as their high-level science products from the STScI MAST, aligned to the \textit{JWST} imaging and registered using SWARP \citep{Bertin2010}. The exception is for CEERS, where \textit{HST} EGS imaging in F606W and F814W was provided in a data release by the CEERS team (see \citealt{Koekemoer2011}).

The global 5$\sigma$ limiting magnitudes for each of the different surveys and in each filter can be found in Table \ref{table:depths}. These global depths were measured in $0.35^{\prime\prime}-$diameter apertures and corrected to total assuming a point-source correction, following the method outlined in \cite{Donnan2023}. Note that the global depth in a given survey is subject to potentially significant spatial variations in local depth, owing to differences in exposure time, source-crowding (particularly for the Quintet, Cartwheel and cluster fields) or sensitivities of the NIRCam modules. We note that these quoted depths are somewhat shallower than those of other studies who use smaller apertures, e.g. $0.2^{\prime\prime}-$diameter such as \citet{Bagley2023} in the data release for CEERS. While the adoption of larger apertures is a more conservative approach, it is still instructive to verify that our reductions are of a similar quality and depth to those used in other studies. We therefore tested that using an equivalent aperture size would yield similarly deep limiting magnitudes. Our resulting depths were consistent to within $\simeq0.1$ magnitudes of the quoted depths by \citet{Bagley2023}. We also include the effective survey area for each of the fields, accounting for de-lensing where applicable.

\begin{table*}
\begin{tabular}{lcccccccccccccc}
\hline
Filter& SM0723 & Quintet & C'wheel & W0137 & NEP-TDF & J1235 & M0647 & R2129 & GLASS & DDT & UNCOVER & CEERS\\
\hline
F435W & 26.7 & - & - & 26.6 & - & - & 26.7 & 26.8 & - & - & 28.2 & 27.8 \\
F606W & 27.7 & - & - & 27.4 & - & - & 27.2 & 27.2 & - & -& 28.6 & 28.4 \\
F814W & 26.8 & - & - & 26.8 & - & - & 27.2 & 26.7 & - & -& 28.4 & 28.0 \\
F070W & - & - & - & - & - & 27.9 & - & - & -& - & -& - \\
F090W & 28.5 & 26.9 & 27.3 & 27.7 & 28.1  & 28.4 & - & -& 28.8 & -& - & - \\
F115W & - & - & - & 27.9 & 28.2 &  28.5 & 27.7 & 27.2 & 28.9 & 28.5 & 28.6 & 28.5 \\
F150W & 28.6 & 27.0 & 27.3 & 28.0 & 28.0  & 28.4 & 27.8 & 27.8 & 28.5 &  28.6 & 28.6 & 28.4 \\
F200W & 28.7 & 27.3 & 27.6 & 28.1 & 28.3  & 28.8 & 28.2 & 27.6 & 28.8 & 28.7 & 28.8 & 28.6 \\
F277W & 28.7 & 27.7 & 27.9 & 28.3 & 28.7  & 28.9  & 28.3 & 27.9 & 28.9 & 28.7 & 28.8 & 28.6 \\
F300M & - & - & - & - & - & 28.6 &  - & - & -& - & - & - \\
F356W & 28.7 & 27.8 & 27.9 & 28.2 & 28.5  & 28.9 & 28.3 & 28.3 & 28.9 & 28.9  & 28.9 & 28.5 \\
F410M & - & - & - & 27.9 & 28.1 & -& - & -& - & -  & 28.5 & 28.0 \\
F444W & 28.4 & 27.4 & 27.8 & 27.9 & 28.3 &  28.3 & 28.0 & 27.7 & 28.8 & 28.3 & 28.5 & 28.2 \\
\hline
Area (arcmin$^{2}$) & \phantom{o}6.4 & 40.9 & \phantom{o}4.2 & \phantom{o}5.9 & 10.5 & \phantom{o}9.9 & \phantom{o}5.1 & \phantom{o}3.1 & 10.1 & \phantom{o}6.6 & 12.2 & 95.0\\
\hline
\end{tabular}
\caption{The $5\sigma$ global limiting magnitudes for each of the fields analysed in this study. These have been measured in $0.35^{\prime\prime}-$diameter apertures on the PSF-homogenized imaging, and corrected to total assuming a point-source correction. Note that for some fields e.g. the clusters and Quintet, these global values are subject to significant variations (as high as $>0.5$ mag) across the field of view owing to crowding and enhanced background, as illustrated in Fig. \ref{fig:depth_map}. We also provide the effective areas in each field, accounting for cluster subtraction and de-lensing where applicable.}
\label{table:depths}
\end{table*}

\begin{figure}
\centering
\includegraphics[scale=1]{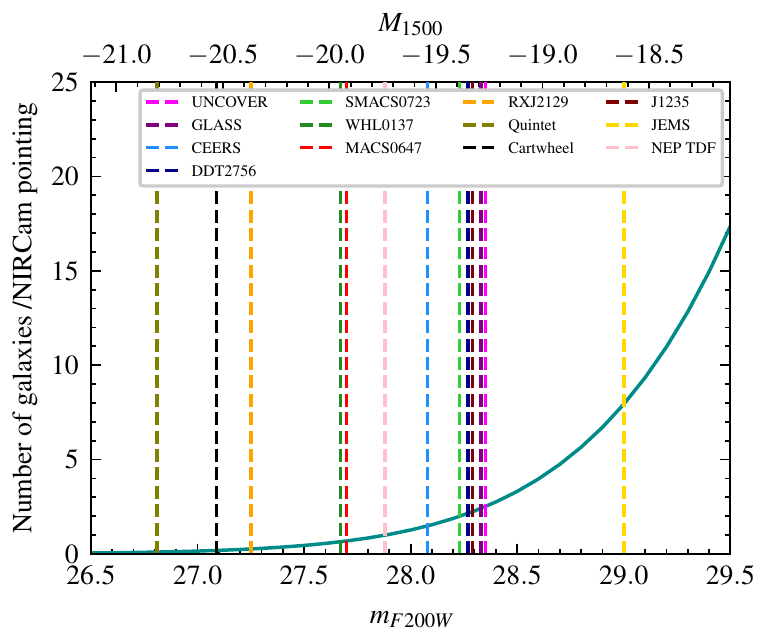}
\caption{An illustration of the distribution of F200W $8\sigma$ global depths across the various surveys utilised in our search for $z \geq 9.5$ galaxies. These depths have been calculated in $0.35^{\prime\prime}-$diameter apertures and corrected to total assuming a point-source correction, following the method in \citet{Donnan2023}. Although we do not perform a new search over JEMS for this study, we utilise the sample as found in \citet{Donnan2023b}, and so we include the F182M depth here. The solid teal line shows the predicted number of galaxies at $z \geq 9.5$ in the area of one NIRCam pointing assuming the luminosity functions from \citet{Donnan2023}. The top $x-$axis shows the absolute UV magnitude for a $z=11$ galaxy with an apparent magnitude given by bottom $x-$axis at $z=11$.}
\label{fig: surveys}
\end{figure}

We illustrate the distribution in depths across the various surveys in Fig. \ref{fig: surveys}. Assuming a flat UV spectral slope (f$_{\nu}$), we derive a rest-frame $M_{1500}$ limiting magnitude for each survey, corresponding to a $z=11$ galaxy at the 8$\sigma$ apparent magnitude limit in the F200W filter. Using the UV luminosity function determined in \citet{Donnan2023}, we also show the predicted average number of $z\sim11$ galaxies found per NIRCam pointing ($\simeq9.6$ sq. arcmin).
\subsubsection{SMACS 0723}

In this study, we cover three Early Release Observations \citep{Pontopiddan2022} surveys: the SMACS 0723 cluster, Stephan's Quintet and the Cartwheel galaxy. All three of these surveys are comprised of imaging in six NIRCam filters spanning $0.9-4.4\,\mu$m: F090W, F150W, F200W, F277W, F356W and F444W.

The first ERO field we will discuss is the SMACS 0723 cluster field (PID 2736, PI Pontoppidan; \citealt{Pontopiddan2022}). This field has already seen extensive study by various groups in search of high-redshift candidates since the July 2022 data release (e.g. \citealt{Atek2023, Adams2023, Donnan2023, Harikane2023, Bouwens2023}). This survey is covered by two NIRCam modules, with one module centred on the cluster field and the other on a ``blank'' parallel field. Due to lensing by the foreground galaxy cluster, which serves to magnify objects in the background, there is also a reduction in the effective area of the search.

To avoid uncertain lensing volume corrections and areas with significantly shallower depths than the global depth, we remove the portion of the image worst affected by intracluster light and enhanced crowding. We then determine the magnification of sources and the remaining effective survey area by utilising the lensing model from \citet{Pascale2022}. Although the raw survey area of the SMACS 0723 field is 11.5 sq. arcmin, after excluding the cluster and accounting for lensing this is reduced to an effective area of 6.4 sq. arcmin. This area is calculated with the assumption that any part of the field-of-view that is not covered by a gravitational lensing map has a magnification $\mu$=1.2, which is what we adopt for all cluster+parallel pair surveys to take into account that the ``blank'' field is still subject to magnification (see Section \ref{Section5}).

Over the cluster portion of the field there is ancillary \textit{HST} data from the RELICS survey (PI Coe; \citealt{Coe2019}), with ACS imaging in F435W, F606W and F814W as well as WFC3/IR imaging in F105W, F125W, F140W and F160W. We include the \textit{HST} ACS data in order to provide wavelength coverage in the optical portion of the spectrum. Although the $5\sigma$ limiting magnitudes of the \textit{HST} ACS data are significantly shallower than the {\it JWST} data, the data still serves to verify any non-detections short-ward of the Lyman break. The $5\sigma$ depths ($0.35^{\prime\prime}-$diameter apertures corrected to total) of the ACS F435W and F814W imaging are approximately 26.8 mag, with the F606W deeper at 27.7 mag.

\subsubsection{Stephan's Quintet}
The Stephan's Quintet field was also imaged as part of the Early Release Observations (PID 2732, PI Pontoppidan; \citealt{Pontopiddan2022}). To date, this field has had one published search for high-redshift galaxies by \citet{Harikane2023}, which yielded one galaxy candidate at $z\simeq12$ and another candidate at $z\simeq16$. This field has the same NIRCam filter set as SMACS 0723, but lacks sufficiently deep ancillary \textit{HST} data. The Quintet field has significant variations in depth across the imaging owing to crowding and enhanced background light. The overall area of the field is approximately 46.7 sq. arcmin, which reduces to an effective area of 40.9 sq. arcmin after removing the Quintet.

\subsubsection{Cartwheel galaxy}
The Cartwheel galaxy, or PGC 2248, was imaged as part of the Early Release Observations (ERO PID 2727, PI Pontoppidan). Once again, the complement of NIRCam filters for this field is the same as for SMACS 0723 and Stephan's Quintet, and there is no useful ancillary \textit{HST} ACS imaging. Similarly to the Quintet field, the depth across the imaging varies significantly, even when masking out the Cartwheel galaxy itself. The field spans $\sim5.7$ sq. arcmin, which reduces to 4.2 sq. arcmin after masking out the Cartwheel galaxy and the two large foreground galaxies to the north east.

\subsubsection{WHL 0137}
The WHL0137 cluster, known for the lensed ``sunrise arc'' galaxy and the $z=6.2$ star candidate ``Earendel'', was imaged as part of Cycle 1 GO PID 2282 (PI Coe; see \citealt{Welch2022, Bradley2023}). Imaging was taken over a single NIRCam pointing, with one module on the cluster and the other on a parallel blank field. The filter set includes the SW channel F090W, F115W, F150W and F200W bands, as well as the LW channel F277W, F356W, F410M and F444W bands. The raw survey area of this field is approximately 10.5 sq. arcmin, which reduces to an effective area of 5.9 sq. arcmin when accounting for the gravitational lensing. This data set has been the subject of a previous search for high-redshift galaxies by \citet{Bradley2023}, which yielded a sample of four $z=8-10$ candidates. Our current study covers the first epoch of observations from July 2022.

As with SMACS 0723, this field has ancillary \textit{HST} RELICS \citep{Coe2019} data over the cluster region, which we use both in SED fitting and to check against any potential low-redshift interlopers in our final sample. The $5\sigma$ depths ($0.35^{\prime\prime}-$diameter apertures, corrected to total) of the ACS F435W and F814W imaging are approximately 26.7 mag, with the F606W imaging significantly deeper at 27.4 mag.

The lensing maps we utilise for WHL 0137 were those prepared for the RELICS survey: Zitrin-LTM (see \citealt{Broadhurst2005,Zitrin2015}, Glafic \citep{Oguri2010} and LENSTOOL \citep{Jullo2009,Kneib2011}.

\subsubsection{MACS 0647}
The MACS 0647 galaxy cluster was previously one of the \textit{HST} CLASH clusters \citep{Postman2012}. The field in which MACS 0647 resides is known to contain a triply-lensed $z=11$ galaxy candidate, MACS0647-JD1 \citep{Coe2013}, which was one of the highest-redshift galaxy candidates in the pre-\textit{JWST} era. MACS 0647 was imaged as part of \textit{JWST} GO proposal 1433 (PI Coe) in order to follow-up this object, with a cluster+parallel pair of NIRCam modules imaged in F115W, F150W, F200W, F277W, F356W and F444W. With this imaging, taken in September 2022, the object was re-analysed and found to be a potential merger system at $z\sim10$ \citep{Hsiao2023}. Most recently, the object was spectroscopically confirmed to lie at $z=10.17$ \citep{Harikane2023b}. 

To complement the \textit{JWST} photometry, we also employ the \textit{HST} ACS optical imaging from the CLASH survey, which covers the cluster module and has a $5\sigma$ depth of $\simeq$27.0 mag. The raw area of the MACS 0647 field was found to be 10.5 sq. arcmin, which reduces to 5.1 sq. arcmin after de-lensing, and the removal of un-useable regions due to the presence of particularly bright stars.

We use the lensing maps that were produced as part of CLASH, i.e the Zitrin-NFW and Zitrin-LTM-Gauss models \citep{Zitrin2015}.

\subsubsection{RXJ 2129}
Another \textit{HST} CLASH cluster, the RXJ 2129 field was imaged in October 2022 as part of Director's Discretionary Time (DDT, PID 2767; PI Kelly). One NIRCam module was imaged in F115W, F150W, F200W, F277W, F356W and F444W, although the exposure times vary significantly between filters, with the F356W imaging being particularly deep. The complement of {\it HST} imaging matches that of MACS 0647. The lensing maps also include Zitrin-LTM-Gauss and Zitrin-NFW but with an additional model from \citet{Caminha2019}. This field was initially selected as a DDT proposal to follow up a supernova at $z=1.5$, but has also yielded a very faint, yet highly magnified ($\mu\sim20$) galaxy confirmed with spectroscopy to be at $z=9.51$ \citep{HWilliams2023}. The raw area of this field is 5.3 sq. arcmin, with a de-lensed effective area of 3.1 sq. arcmin.

\subsubsection{GLASS}
As part of \textit{JWST} PID 1324 (PI Treu; \citealt{Treu2022}; see also \citealt{Paris2023}), a pair of NIRCam pointings were taken in parallel to NIRISS observations of the Abell 2744 cluster field. The first epoch of observations was taken in July 2022, and has been the subject of numerous searches for high-redshift galaxies, including the discovery of an exceptionally bright $z=12$ galaxy candidate GLASS-z12 \citep{Castellano2022, Naidu2022}. The second epoch of observations was taken in November 2022, and has yielded yet further exciting results, including the discovery of an over-density of bright $z=10$ galaxy candidates \citep{Castellano2023}.

We include both epochs of observations in our present study. The NIRCam filter set comprises F090W, F115W, F150W, F200W, F277W, F356W and F444W. One NIRCam module benefits from increased exposure time from repeated observation, and there is an additional two NIRCam modules with significant overlap, producing a total raw search area of 13.3 sq. arcmin.

Some of the GLASS parallel field is covered by the \citet{Furtak2023} lensing model of the Abell 2744 cluster, which was released as part of UNCOVER Data Release 1. Where there is coverage we adopt this lensing map, and set a floor of $\mu$=1.2 for the rest of the area that is not covered. This gives us an effective de-lensed area of 10.1 sq. arcmin.

\subsubsection{North Ecliptic Pole Time Domain Field (NEP TDF)}
The NEP TDF field was imaged in September 2022 as part of the PEARLS survey (PID 2738, PI Windhorst; see \citealt{Windhorst2023}). NIRCam imaging was taken in F090W, F115W, F150W, F200W, F277W, F356W, F410M and F444W. The majority of this data set is proprietary, however one of the pointings in the NEP TDF was made publicly available immediately for use by the community. Although the footprint of the F150W and F356W imaging is larger, we focus on the $\simeq10.5$ sq. arcmin overlap between all of the available NIRCam filters.

\subsubsection{CEERS}
The Cosmic Evolution Early Release Science Survey (CEERS, ERS 1345, PI Finkelstein; see \citealt{Bagley2023} for imaging release) is an early release science program covering approximately 100 sq. arcmin of the CANDELS Extended Groth Strip (EGS) field. This field has been studied extensively since the first epoch of observations was taken in July 2022 (e.g. \citealt{Finkelstein2022, Finkelstein2023, Donnan2023, Bouwens2023}). This study has also already seen the spectroscopic confirmation of several high-redshift galaxies at $z=8-11$, including ``Maisie's galaxy'' \citep{Finkelstein2022, ArrabalHaro2023, ArrabalHaro2023b}.

In this study, we include an analysis of both the first and second epochs of observations, taken in July and December 2022, respectively. \textit{JWST} imaging is available in seven NIRCam bands: F115W, F150W, F200W, F277W, F356W, F410M and F444W. We also employ \textit{HST} ACS imaging to provide additional wavelength coverage in the optical. As noted earlier, a reduction of the \textit{HST} imaging in ACS F606W and F814W has been made publicly available by the CEERS team (HDR1; see \citealt{Koekemoer2011}). Additional F435W imaging was available from the UVCANDELS survey (GO 15647, PI Teplitz; DOI: 10.17909/8s31-f778).

\subsubsection{UNCOVER}
The UNCOVER survey (Cycle 1 2561, PI Labbe; \citealt{Bezanson2022}) is an ultra-deep survey of the Abell 2744 cluster field covering a raw area of $\simeq 29$ sq. arcmin ($\simeq 12$ sq. arcmin after delensing and masking) to a depth in excess of 29 mag (0.35$^{\prime\prime}-$diameter apertures). The NIRCam filter set is F115W, F150W, F200W, F277W, F356W, F410M and F444W. The lensing map used for this data set was the \citet{Furtak2023} model that was publicly released as part of UNCOVER's Data Release 1. We also employ this lensing map for the regions of GLASS and DDT 2756 where there is coverage.

Abell 2744 was imaged with \textit{HST} as one of the Hubble Frontier Fields \citep{Lotz2017}. This survey includes among the deepest \textit{HST} observations in the ACS and WFC3/IR filters, with depths of $28.2-28.6$ mag. Although the ACS observations cover less than half of the UNCOVER survey area, we include the ACS F435W, F606W and F814W data in order to extend our wavelength coverage into the optical.

\subsubsection{DDT 2756: Abell2744}
The Abell 2744 cluster was imaged with NIRCam as part of DDT program 2756 (PI Chen), a program designed to follow-up a $z=3.47$ supernova. Two epochs of observations in six NIRCam bands (F115W, F150W, F200W, F277W, F356W and F444W) were taken in October 2022 and December 2022. Although the Abell 2744 module imaging was included in the UNCOVER reduction, each epoch includes an additional parallel NIRCam module. These two partially overlap, but together provide another 6.6 sq. arcmin of effective search area. This field was also found by \citet{Castellano2023} to contain a very bright $z\simeq10$ galaxy candidate, with reported $M_{UV}\simeq-21.60$. Ultimately, the object has been spectroscopically confirmed to lie at $z=9.31$, and is likely an interacting system \citep{Boyett2023}. That being said, an independent search over both epochs of data for this survey is clearly desirable to discover if there are other exceptionally bright candidates at similarly extreme redshifts.

\subsubsection{PID 1063 Nircam Flats: J1235}
As part of NIRCam commissioning (PID 1063, PI Sunnquist), exceptionally deep imaging was taken of a field centred at RA=12:35:48, Dec$=+$04:55:45. The filter set included is also more extensive than the other \textit{JWST} surveys: F070W, F090W, F115W, F150W, F200W, F277W, F300M, F356W and F444W. The coverage across this rich filter set is not homogeneous, and so we restrict our analysis to an area of $\simeq9.9$ sq. arcmin covered by all of the available filters. This field was recently studied as part of a search for quiescent galaxies by \citet{Valentino2023}.

\subsubsection{JEMS}
Deep medium band observations of the HUDF were taken as part of the JEMS survey (ID 1963; PI Williams; see \citealt{Williams2023udf}). This included SW F182M and F210M imaging along with LW imaging in F430M, F460M and F480M. The total $5\sigma$ global depths are $\simeq29$ mag for the SW channel filters and $\simeq28.5$ mag for the LW channel filters. 

Using this data set, \citet{Donnan2023b} uncovered a sample of six high-redshift candidates, including three spectroscopically confirmed $z>10$ galaxies \citep{Curtis-Lake2023, Robertson2023}. Among these is UDFj-39546284, a $z\simeq12$ galaxy first discovered a decade previously with \textit{HST} imaging \citep{Bouwens2011, Ellis2013}. Although we do not search the JEMS data set in this study, we include the six candidates found in \citet{Donnan2023b} in our LF analysis in Section \ref{Section5}.

\section{Sample Selection}
\subsection{Catalogue Construction}
For each of the fields, we construct multi-wavelength catalogues using \textsc{Source-Extractor} \citep{Bertin1996} in dual-image mode. As our study is focused on selecting $9.5<z<12.5$ galaxies, the position of the Lyman break will move through the F150W filter as we probe to higher redshift, with the IGM absorption having attenuated half of the flux in the F150W band by $z=11$. Hence, in order to facilitate the detection of galaxies beyond a Lyman break of 1.5 $\mu$m, we use the F200W imaging as our primary detection band. We create additional catalogues using each of the broadband LW filters (F277W, F356W, F444W) as detection images, in order to supplement this first catalogue with any sources that may have been missed. For the three fields lacking F115W coverage (SMACS 0723, Cartwheel, Quintet), we cannot search as effectively for $z<11$ galaxies and a F150W detection image would not add significant value. Hence for consistency across all of the fields, we do not include F150W detection images in this study. We also adjust our selection criteria in these three fields to account for the reduced filter coverage (see Section \ref{SelectionSection}).

Prior to any aperture photometry, we homogenized the PSF of all of the NIRCam images to match that of the F444W image, in order to minimize any colour systematics in subsequent analyses. Aperture photometry was performed on these PSF-homogenized images with fluxes measured in $0.35^{\prime\prime}-$diameter apertures. This aperture size is designed to contain close to 75\% of point-source flux for the F444W PSF. We note that this choice of aperture size is reasonable given the sizes of $z=9-10$ galaxies found previously in the literature, e.g. \citet{Morishita2018} and \cite{Naidu2022}. Photometric uncertainties were calculated object-by-object by taking the median absolute deviation of the nearest 150-200 blank-sky 0.35$^{\prime\prime}-$diameter apertures, and scaling to $\sigma$ by multiplying by 1.4826, following the method by \citet{McLeod2021}. Aperture fluxes across all filters were then corrected (typically at the $\simeq$1-2\% level) to 75\% of point-source total to take into account any remaining inhomogeneities in the curves of growth between filters after PSF homogenization.

For each of our initial catalogues, we required a $5\sigma$ detection in the detection image, as well as an additional $3\sigma$ detection in at least one other band to reduce the number of spurious detections.

\subsection{SED-fitting and photometric redshifts}
We performed SED fitting on all of the sources within our initial catalogues using the SED fitting code \textsc{LePhare} \citep{Arnouts2011}. The SED library that we used was \citet{BC03} (BC03), using a \citet{Chabrier2003} initial mass function, a \citet{Calzetti2000} dust attenuation law and the IGM absorption prescription from \citet{Madau1995}. These SED templates included declining star-formation histories with various $\tau$ values from 0.1 to 15 Gyr, and metallicities of 0.2\,Z$_{\odot}$ and Z$_{\odot}$. We included emission lines, and the $A_{V}$ was allowed to range from 0 to as high as 6 to clean the sample of any extremely reddened low-redshift interlopers masquerading as high-redshift galaxies.

We measured rest-frame $M_{1500}$ magnitudes using a 100\,\AA-wide tophat filter on the best-fitting SED. Given the remarkably high resolution of \textit{JWST}'s NIRCam imaging, even the highest-redshift galaxies are often not well approximated by point-sources, and so require larger corrections to total than point-source corrections. Hence, rather than simply scaling to point-source total flux, we correct the $M_{1500}$ magnitudes by scaling to the \textsc{FLUX\_AUTO} \citep{Kron1980} value measured by \textsc{Source-Extractor}, and we add a further 10\% correction to account for the light outside of the Kron aperture. This further correction is motivated by an analysis performed on stacks of our extended source $z\geq9$ candidates (see Section \ref{candidates_section}), where we found that \textsc{FLUX\_AUTO} under-estimated the flux versus a large $1.2^{\prime\prime}-$diameter aperture by $\simeq10\%$.
\subsection{Selection of high-redshift candidates}
\label{SelectionSection}
We initially selected candidates with a photometric redshift solution $z_{\mathrm{phot}}\geq8.5$ and a goodness of fit $\chi^{2}_{\nu}\leq10$. Only candidates considered ``robust'', with $\Delta\chi^{2}\geq4$ between the primary and secondary photo-z solutions, were retained for the final sample. We also further refined our sample selection to require a $\geq8\sigma$ detection in at least one of the F150W, F200W and F277W images, to ensure that we were only analysing high signal-to-noise objects. Due to the IGM absorption of any flux short-ward of the Lyman break at $z\leq9.5$, we further required a non-detection ($\leq2\sigma$) in the F090W band, as well as in F115W and the \textit{HST} ACS imaging if available. The condition of F115W$<$2$\sigma$ precludes any complete statistical investigation of the $8.5<z<9.5$ population, given that a $z\simeq9.0$ galaxy can still be detectable in F115W. Consequently, we confine our later LF analysis to $z>9.5$, but include any $8.5<z<9.5$ candidates in our tabulated sample.

As a validation exercise, we double-checked any high-redshift candidates did not present a preferred low-redshift solution when performing SED fitting using the code \textsc{EAZY} \citep{Brammer2008}, including the \textsc{Pegase} \citep{Fioc1999} templates. The \textsc{EAZY} photometric redshift solution was determined allowing the templates to vary over the redshift range $0<z<25$.

As well as performing the flux cuts to remove any significant detections shortward of the break, we inspected the F090W, F115W and ACS imaging, in order to verify that our candidates do not have any marginal detections.

\begin{figure*}
\begin{tabular}{c|c}
\includegraphics[width=\columnwidth]{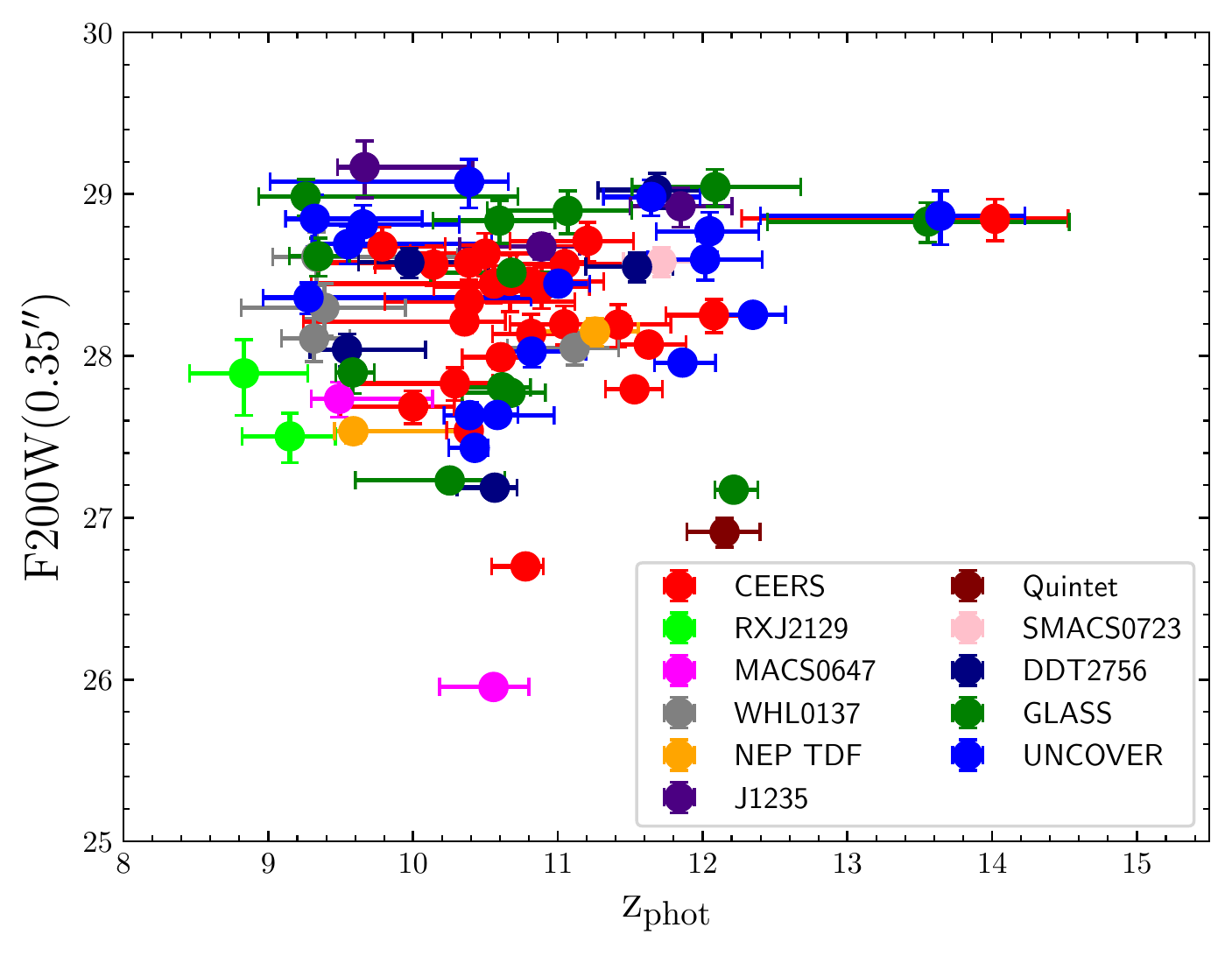}     & 
\includegraphics[width=\columnwidth]{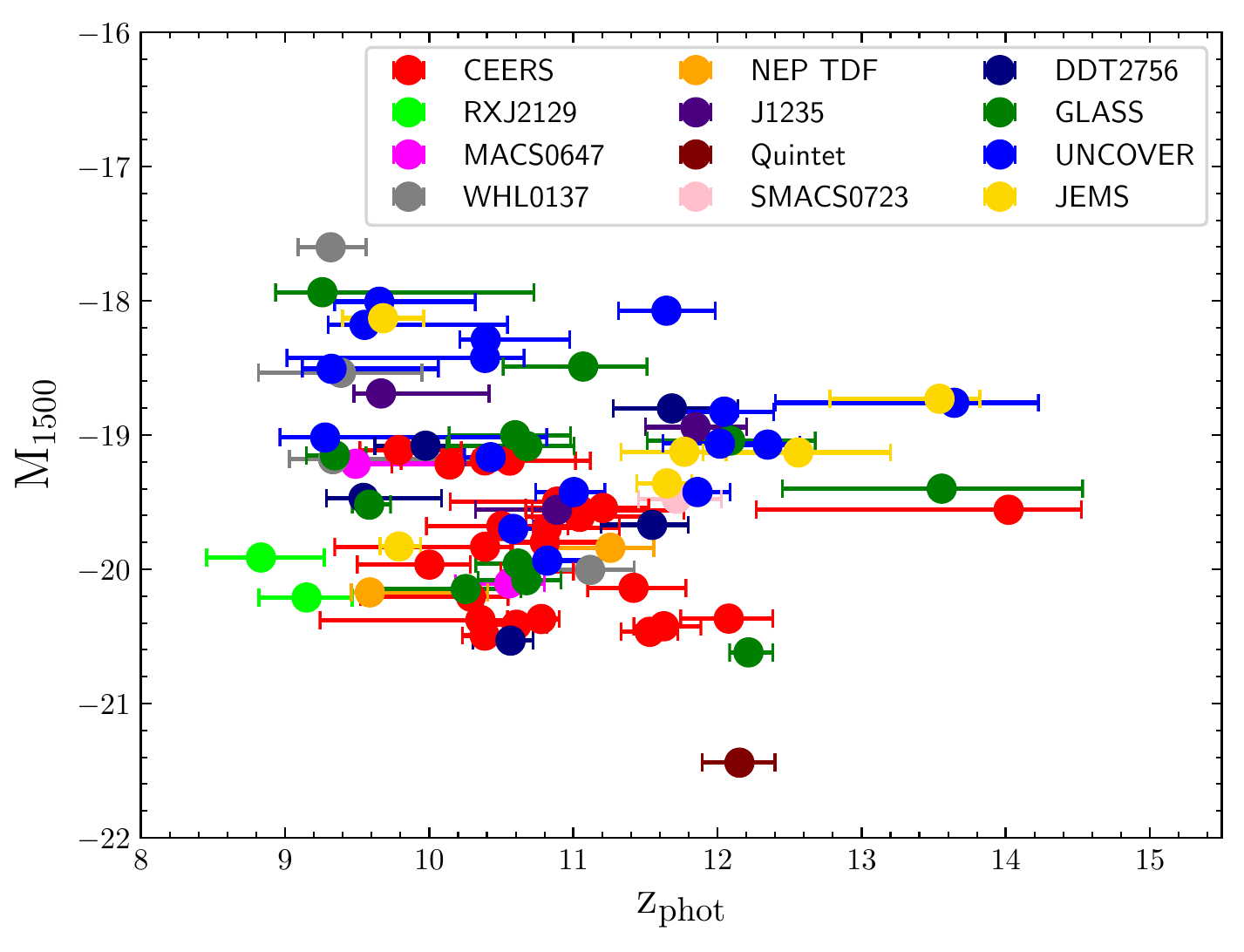} \\ 
\end{tabular}
\caption{The left-hand panel shows the distribution of apparent F200W magnitude (measured in a 0.35$^{\prime\prime}-$diameter aperture) with photometric redshift for our selected sample of high-redshift candidates. The right-hand panel shows the corresponding de-lensed rest-frame $M_{1500}$ magnitude (corrected to total) distribution with photometric redshift, including the additional candidates from the \citet{Donnan2023b} study of JEMS.}
\label{fig: Dist}
\end{figure*}

The Cartwheel, SMACS 0723 and Quintet fields lack the F115W filter, and this creates additional challenges in determining the redshift of any candidates in those fields. As the F090W transmission ends at 1.0 $\mu$m, any F090W-dropout followed by a significant F150W detection can essentially lie anywhere between $z\sim7.2$ and $z\sim12.0$, assuming it is not a low-redshift interloper. This issue is mitigated where there is a F115W non-detection, which helps narrow down the redshift solution to $z\gtrsim9$. We therefore included an extra criterion for these fields, whereby we required either a non-detection ($\leq2\sigma$) in F150W, or a red F150W-F200W$\geq 0.75$ colour. Although this shifts our selection window for these three fields to above z$\simeq$11.5, these criteria prevent the final sample from being contaminated by z$\simeq7-8$ interlopers.

\section{High-redshift candidates}
\label{candidates_section}
\begin{figure*}
\begin{tabular}{cc}
\includegraphics[width=0.9\columnwidth]{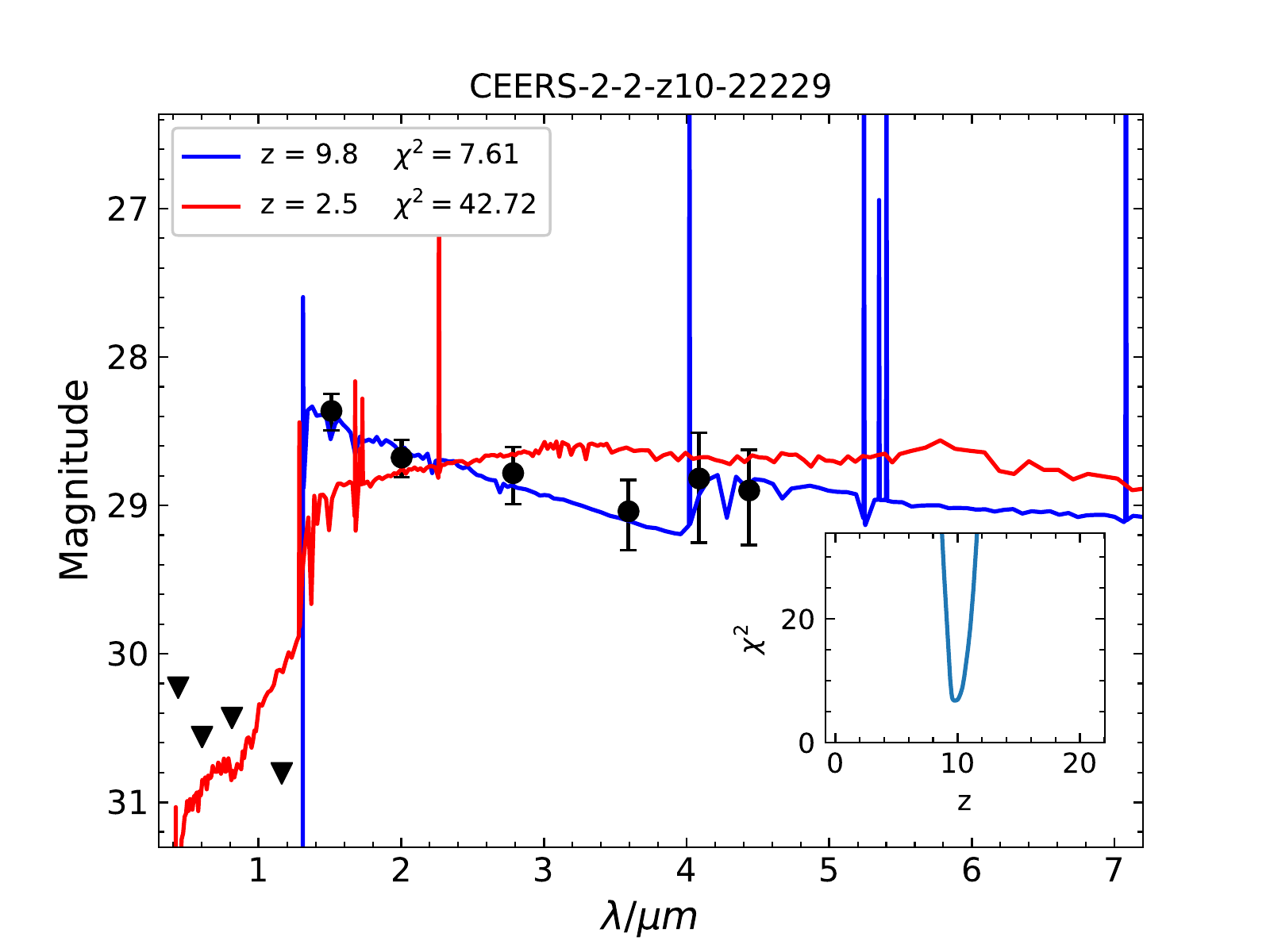} &
\includegraphics[width=0.9\columnwidth]{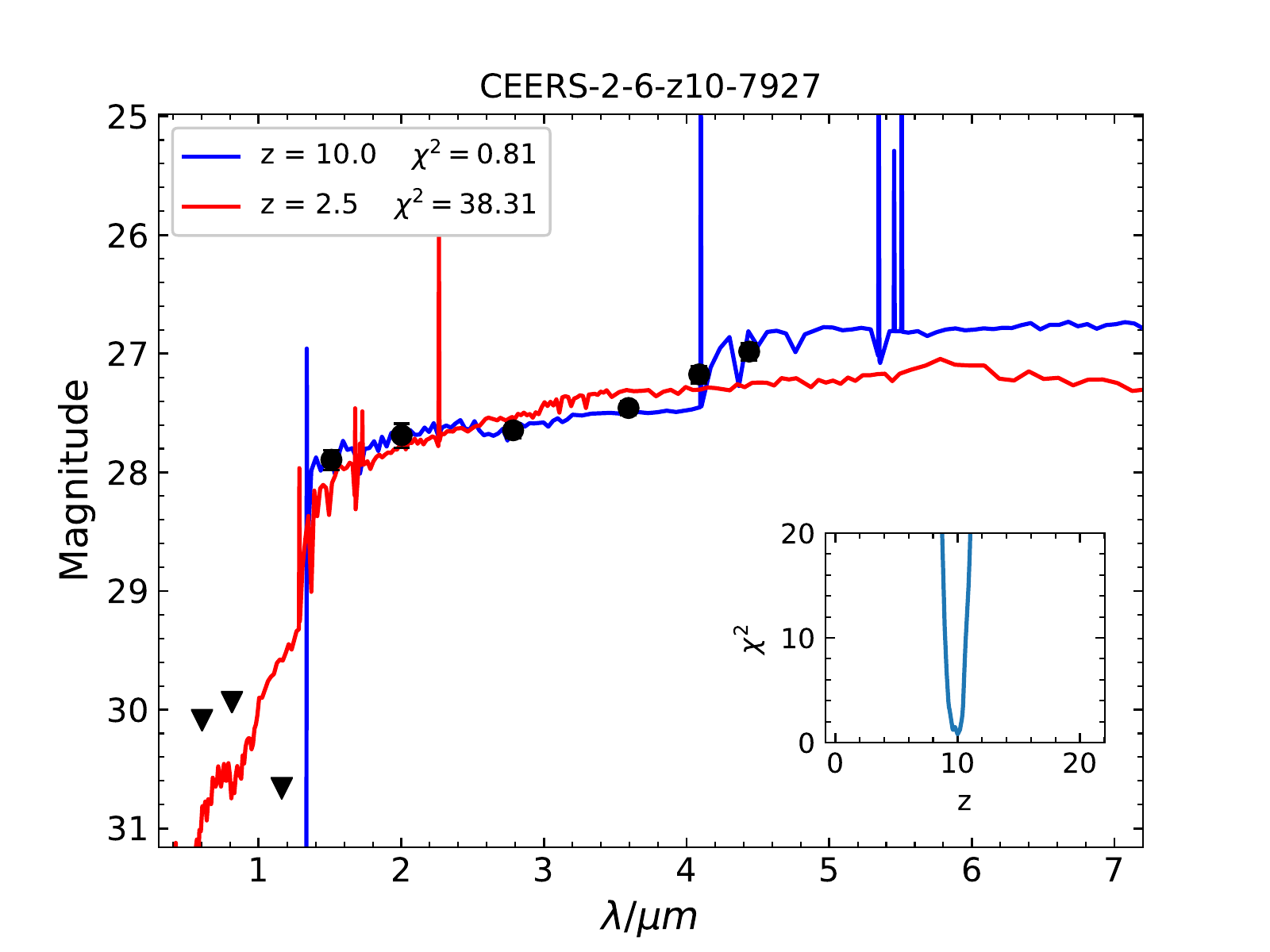} \\
\includegraphics[width=0.9\columnwidth]{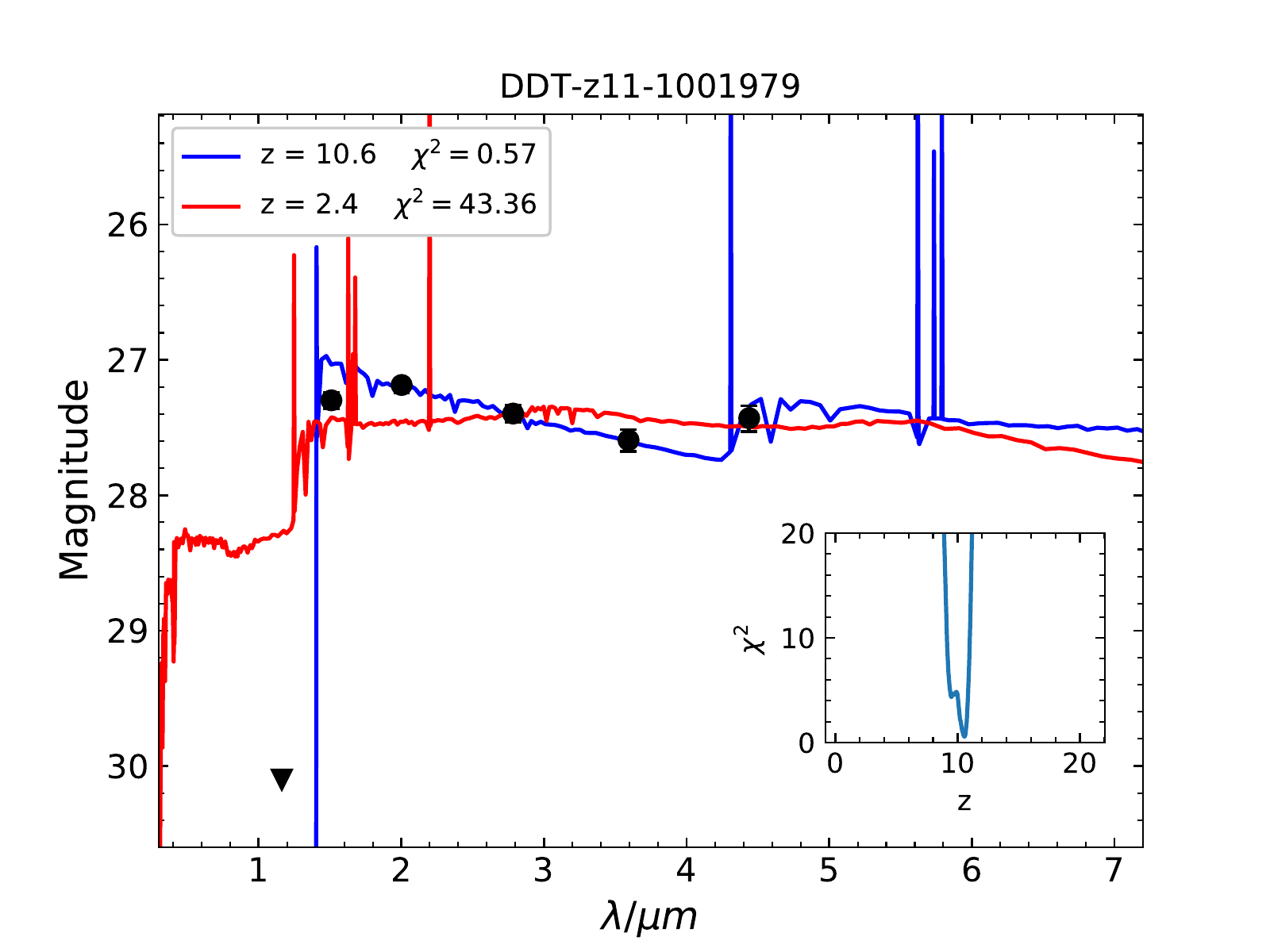} &
\includegraphics[width=0.9\columnwidth]{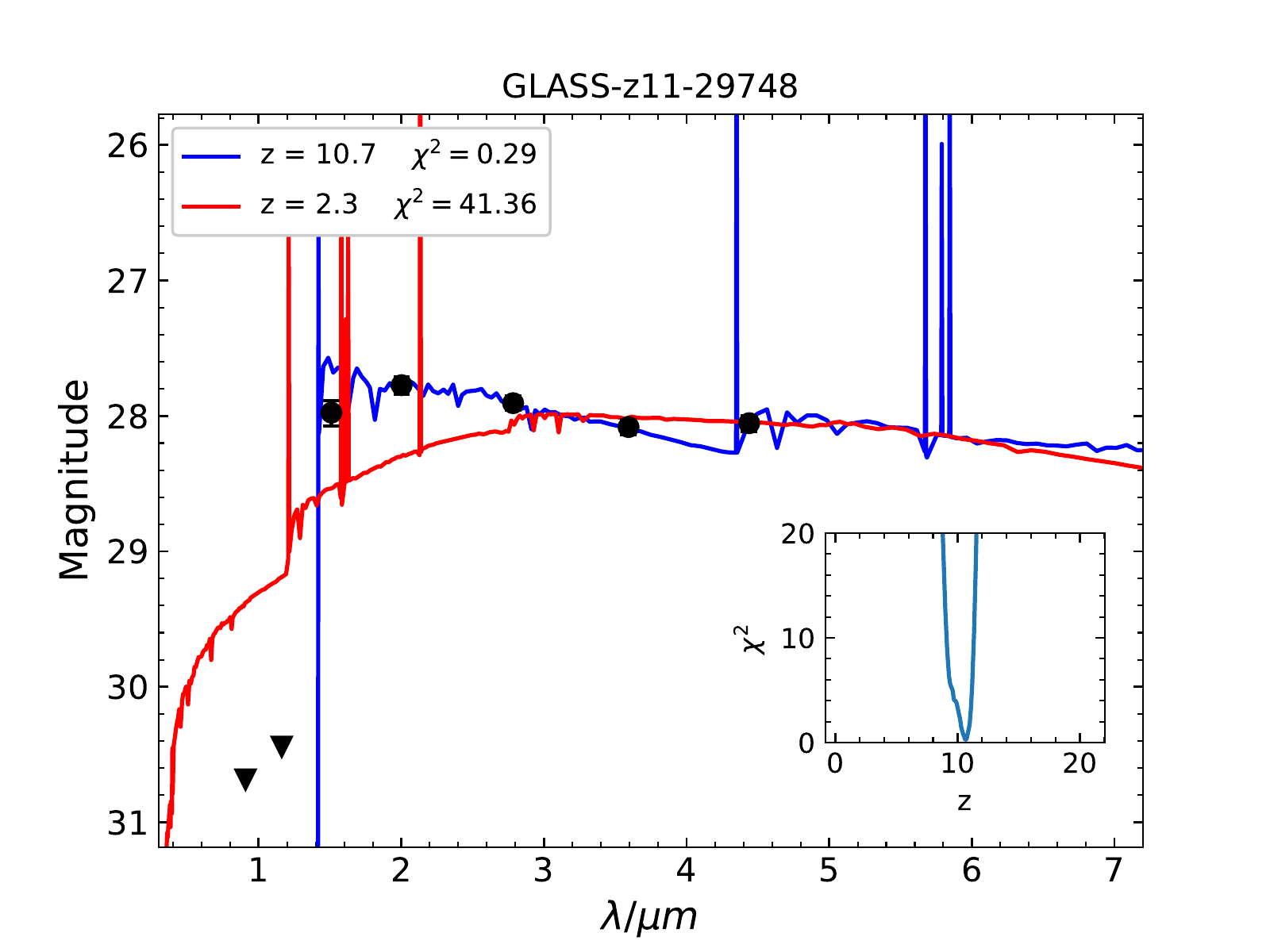} \\
\includegraphics[width=0.9\columnwidth]{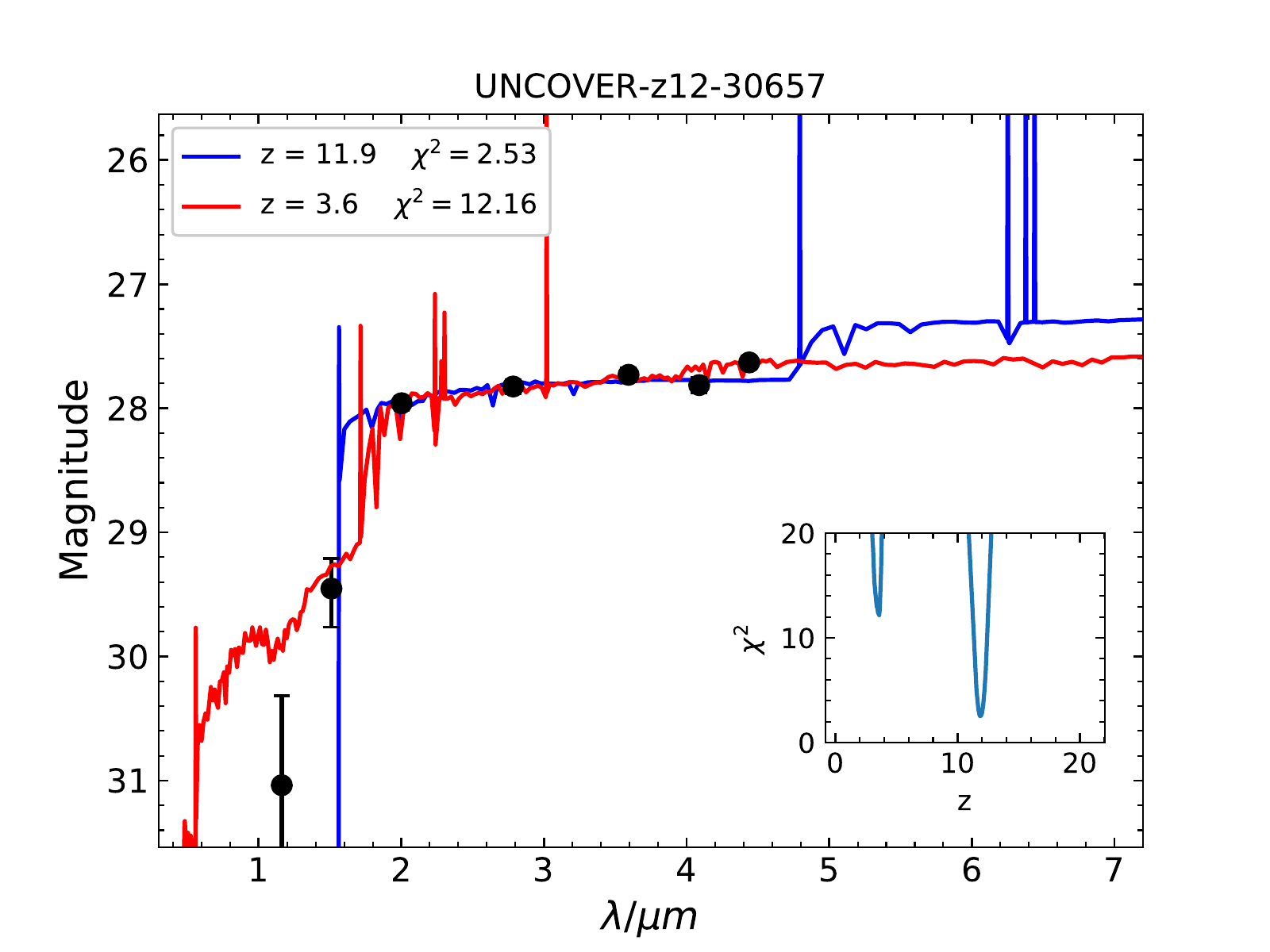} &
\includegraphics[width=0.9\columnwidth]{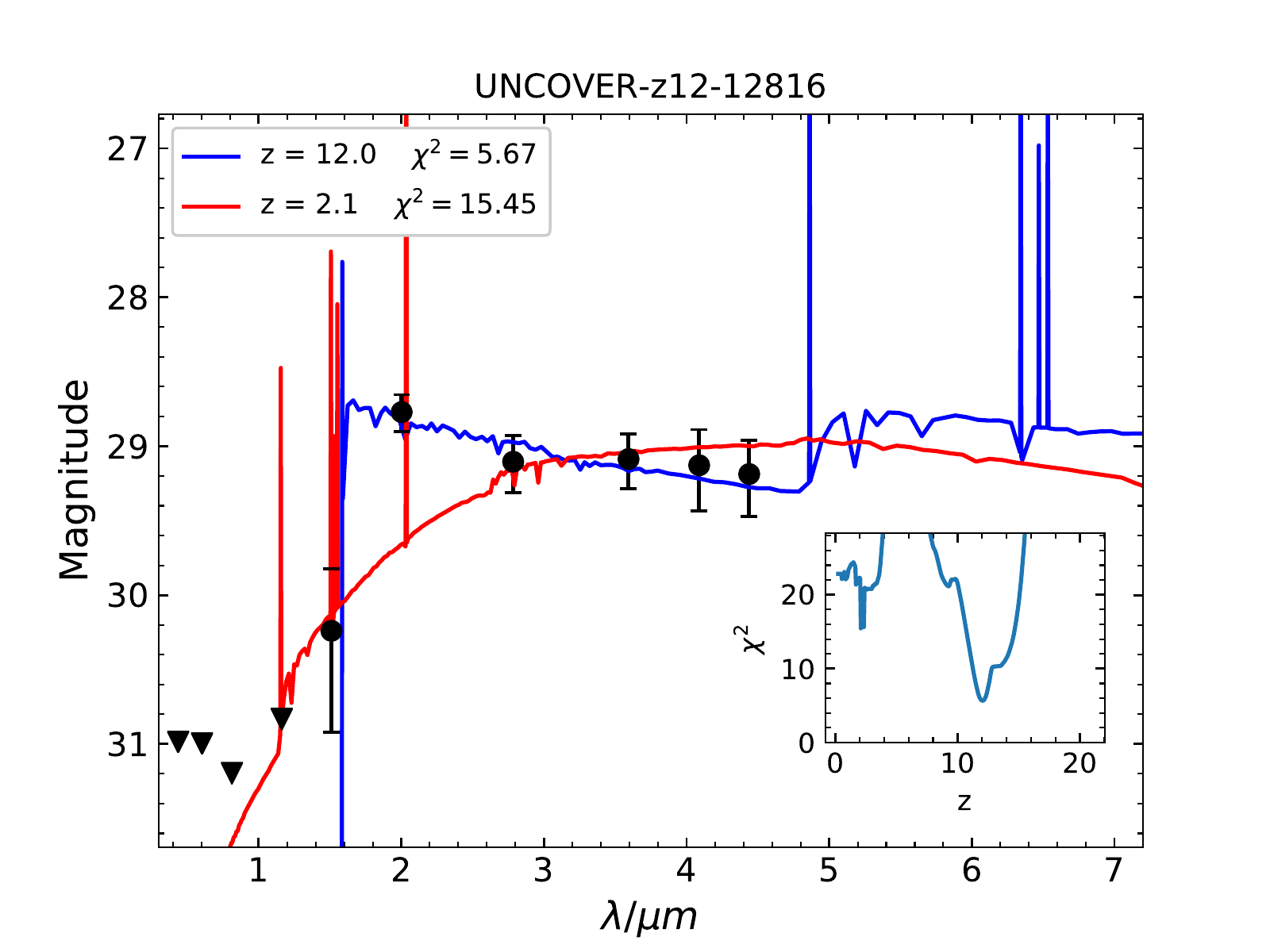}
\end{tabular}
\caption{Examples of \textsc{LePhare} SED plots for high-redshift candidates selected in this study. We include both the primary high-redshift solution (blue line), as well as the competing lower-redshift secondary solution (red line). Each plot includes an inset of the $\chi^{2}$ distribution with redshift for the SED fitting. Downward arrows in the plot denote the 1$\sigma$ upper limit for non-detections.}
\label{fig:SED_plots}
\end{figure*}

\begin{figure*}
\includegraphics[width=2\columnwidth]{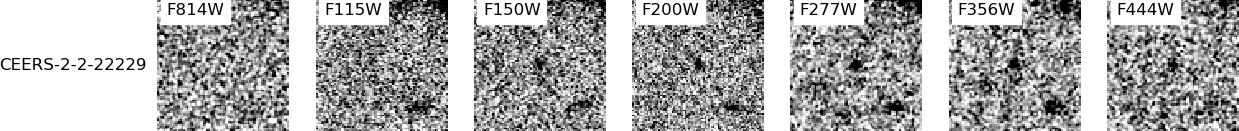} \\[2ex]
\includegraphics[width=2\columnwidth]{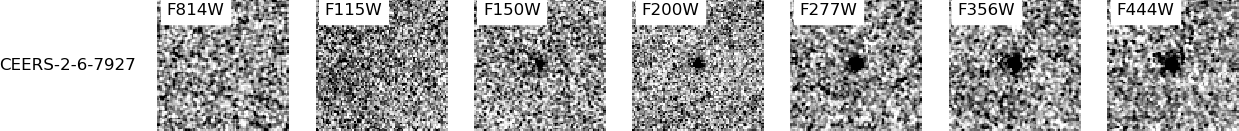} \\ [2ex]
\includegraphics[width=2\columnwidth]{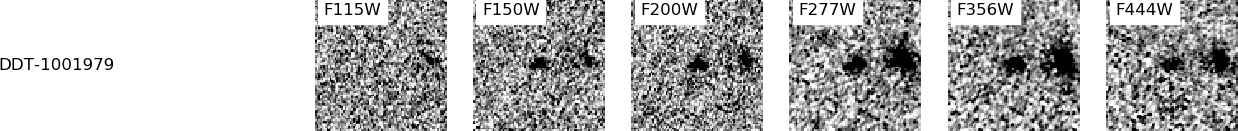} \\ [2ex]
\includegraphics[width=2\columnwidth]{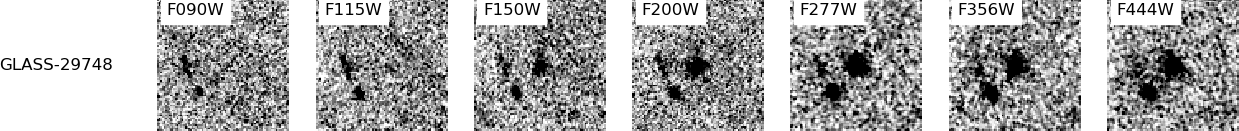} \\ [2ex]
\includegraphics[width=2\columnwidth]{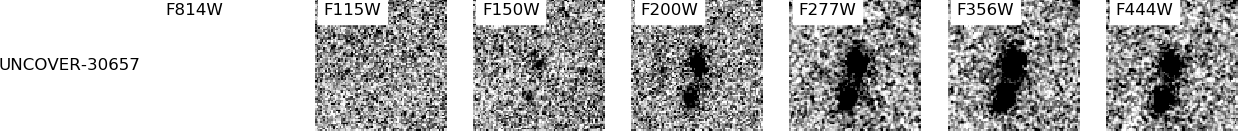} \\ [2ex]
\includegraphics[width=2\columnwidth]{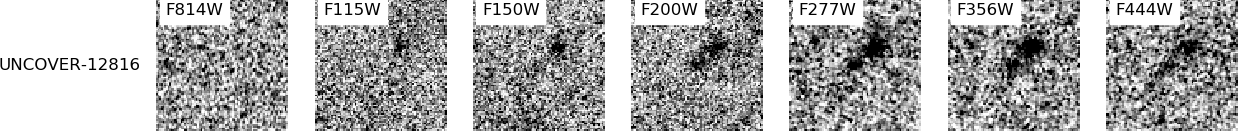} \\ [2ex]
\caption{Postage stamps for each of the candidates shown in Fig. \ref{fig:SED_plots}. The shortest wavelength column is NIRCam F090W or \textit{HST} F814W, depending on availability, followed by each of the broadband NIRCam filters from F115W to F444W. Note that in some surveys lacking ACS ancillary data, e.g. DDT-2756, our shortest wavelength band is F115W. The greyscale has been set to the median background flux $\pm$2$\sigma$, after excluding flux contributions from sources defined by a segmentation map.}
\label{fig: Stamps}
\end{figure*}

In Tables \ref{tab: candidates1}-\ref{tab: candidates2} we list basic properties for the candidates uncovered across the various fields. We also provide a colour-coded illustration of the distribution of intrinsic $M_{1500}$ with redshift, and the distribution of F200W 0.35$^{\prime\prime}-$diameter aperture magnitude with redshift for our final sample in Fig. \ref{fig: Dist}. In Fig. \ref{fig:SED_plots} and Fig. \ref{fig: Stamps} we present some example SEDs and postage stamps for a sub-set of the final candidates. 
\begin{table*}
\begin{tabular}{lcccccc}
\hline
ID & RA & Dec & $\rm z_{phot}$ & $M_{1500}$ & $\rm \mu$ & References\\
\hline
CEERS-NE-z10-3069854 & 14:20:03.35 & $+$53:00:24.37 & $10.4^{+0.2}_{-1.0}$ & $-$19.83 & - & -\\[1ex]
CEERS-2-2-z10-22229 & 14:20:03.68 & $+$52:54:49.37 & $\phantom{o}9.8^{+0.4}_{-0.3}$ & $-$19.11 & - & -\\[1ex]
CEERS-2-3-z10-6303 & 14:19:38.56 & $+$52:52:33.11 & $10.4^{+0.3}_{-1.1}$ & $-$20.38 & - & -\\[1ex]
CEERS-2-5-z10-9617 & 14:18:55.80 & $+$52:45:29.06 & $10.3^{+0.3}_{-0.8}$ & $-$20.20 & - & Ha23\\[1ex]
CEERS-2-5-z10-17402 & 14:19:10.56 & $+$52:46:45.07 & $10.4^{+0.7}_{-0.6}$ & $-$19.19 & - & -\\[1ex]
CEERS-2-6-z10-3530 & 14:19:14.85 & $+$52:44:13.63 & $10.4^{+0.1}_{-0.2}$ & $-$20.49 & - & Ha23\\[1ex]
CEERS-2-6-z10-17416 & 14:19:31.50 & $+$52:46:22.66 & $10.1^{+0.4}_{-0.4}$ & $-$19.22 & - & -\\[1ex]
CEERS-2-6-z10-7927 & 14:19:16.11 & $+$52:44:54.05 & $10.0^{+0.3}_{-0.5}$ & $-$19.97 & - & -\\[1ex]
DDT2756-z10-1010177 & 00:14:35.54 & $-$30:25:35.80 & $10.0^{+0.5}_{-0.4}$ & $-$19.08 & 1.30 & -\\[1ex]
DDT2756-z10-1010612$^{\dagger}$ & 00:14:28.17 & $-$30:25:31.69 & $\phantom{o}9.5^{+0.5}_{-0.3}$ & $-$19.47 & 1.65 & C23\\[1ex]
GLASS-z10-3283 & 00:14:02.87 & $-$30:22:18.71 & $10.3^{+0.4}_{-0.7}$ & $-$20.15 & $\phantom{o}1.68$\phantom{o} & C23, N22, D23, H23, B23\\[1ex]
GLASS-z10-15864 & 00:13:54.90 & $-$30:20:43.95 & $\phantom{o}9.6^{+0.1}_{-0.1}$ & $-$19.52 & \phantom{o}1.20$^{\star}$ & C23\\[1ex]
GLASS-z9-32411 & 00:13:49.13 & $-$30:19:00.85 & $\phantom{o}9.3^{+0.2}_{-0.2}$ & $-$19.15 & \phantom{o}1.20$^{\star}$ & -\\[1ex]
GLASS-z9-9023 & 00:14:03.99 & $-$30:21:33.00 & $\phantom{o}9.3^{+1.5}_{-0.3}$ & $-$17.94 & $\phantom{o}1.65$\phantom{o} & -\\[1ex]
J1235-z10-1046989 & 12:35:52.06 & $+$04:55:03.82 & $\phantom{o}9.7^{+0.7}_{-0.2}$ & $-$18.69 & - & -\\[1ex]
MACS0647-z9-20158 & 06:47:36.95 & $+$70:14:34.69 & $\phantom{o}9.5^{+0.6}_{-0.2}$ & $-$19.21 & $\phantom{o}2.75$\phantom{o} & - \\[1ex]
NEPTDF-z10-1042051 & 17:23:00.91 & $+$65:49:25.20 & $\phantom{o}9.6^{+0.8}_{-0.1}$ & $-$20.17 & - & -\\[1ex]
RXJ2129-z9-16261 & 21:29:38.97 & $+$00:06:50.84 & $\phantom{o}9.1^{+0.3}_{-0.3}$ & $-$20.21 & $\phantom{o}1.24$\phantom{o} & -\\[1ex]
RXJ2129-z9-3021681 & 21:29:40.45 & $+$00:06:43.45 & $\phantom{o}8.8^{+0.4}_{-0.4}$ & $-$19.91 & $\phantom{o}1.29$\phantom{o} & -\\[1ex]
UNCOVER-z9-3658 & 00:14:20.82 & $-$30:25:23.25 & $\phantom{o}9.3^{+0.7}_{-0.2}$ & $-$18.51 & $\phantom{o}2.10$\phantom{o} & -\\[1ex]
UNCOVER-z10-3892 & 00:14:24.69 & $-$30:25:20.55 & $\phantom{o}9.7^{+0.7}_{-0.3}$ & $-$18.01 & $\phantom{o}2.21$\phantom{o} & -\\[1ex]
UNCOVER-z10-23585 & 00:14:29.82 & $-$30:23:20.69 & $10.4^{+0.3}_{-1.4}$ & $-$18.43 & $\phantom{o}1.47$\phantom{o} & -\\[1ex]
UNCOVER-z10-15452 & 00:14:22.79 & $-$30:24:02.56 & $10.4^{+0.6}_{-0.2}$ & $-$18.29 & 10.74\phantom{o} & Z14 \\[1ex]
UNCOVER-z10-32757 & 00:14:16.09 & $-$30:22:40.16 & $10.4^{+0.1}_{-0.2}$ & $-$19.17 & $\phantom{o}3.72$\phantom{o} & C23\\[1ex]
UNCOVER-z10-47904 & 00:14:23.90 & $-$30:21:37.14 & $\phantom{o}9.6^{+1.0}_{-0.3}$ & $-$18.18 & $\phantom{o}1.62$\phantom{o} & -\\[1ex]
UNCOVER-z9-3061884 & 00:14:05.50 & $-$30:21:40.34 & $\phantom{o}9.3^{+1.5}_{-0.3}$ & $-$19.02 & $\phantom{o}1.87$\phantom{o} & -\\[1ex]
WHL0137-10187 & 01:37:25.74 & $-$08:26:38.21 & $\phantom{o}9.3^{+0.2}_{-0.2}$ & $-$17.60 & $\phantom{o}7.95$\phantom{o} & -\\[1ex]
WHL0137-2004877 & 01:37:22.12 & $-$08:28:00.71 & $\phantom{o}9.4^{+0.6}_{-0.6}$ & $-$18.54 & $\phantom{o}2.86$\phantom{o} & - \\[1ex]
WHL0137-21722 & 01:37:19.28 & $-$08:25:25.64 & $\phantom{o}9.3^{+1.0}_{-0.3}$ & $-$19.18 & \phantom{o}1.20$^{\star}$ & -\\[1ex]
\hline
\end{tabular}
\caption{Our final sample of $8.5<z<10.5$ galaxies across all of the fields studied. We include the photometric redshift and intrinsic rest-frame UV magnitude $M_{1500}$, as well as the magnification $\mu$ if applicable. We also include a references column if the candidate has appeared elsewhere (Z14=\citealt{Zitrin2014}, C23=\citealt{Castellano2022,Castellano2023}, N22=\citealt{Naidu2022}, B23=\citealt{Bouwens2023}, D23=\citealt{Donnan2023}, F23=\citealt{Finkelstein2023}, H23=\citealt{Harikane2023}, Ha23=\citealt{ArrabalHaro2023}).\\$^{\star}$ Magnification is fixed to a floor of 1.2 due to a lack of lensing map coverage.\\$^{\dagger}$ DDT2756-z10-1010612 is the north-east component of what is believed to be a merging system, first found by \citet{Castellano2023} and confirmed with spectroscopy to be at $z=9.31$ \citep{Boyett2023}.}
\label{tab: candidates1}
\end{table*}

\begin{table*}
\begin{tabular}{lcccccccc}
\hline
ID & RA & Dec & $\rm z_{phot}$ & $M_{1500}$ & $\rm \mu$ & References \\
\hline
CEERS-NE-z11-17070 & 14:19:37.59 & $+$52:56:43.83 & $11.0^{+0.5}_{-0.4}$ & $-$19.61 & - & F23, D23\\[1ex]
CEERS-NE-z11-39491 & 14:20:10.56 & $+$52:59:39.52 & $10.6^{+0.5}_{-1.3}$ & $-$19.19 & - & F23, D23\\[1ex]
CEERS-NE-z11-2543 & 14:19:41.47 & $+$52:54:41.51 & $11.0^{+0.7}_{-0.3}$ & $-$19.56 & - & F23, D23\\[1ex]
CEERS-NE-z12-15760 & 14:19:46.35 & $+$52:56:32.82 & $11.6^{+0.3}_{-0.2}$ & $-$20.42 & - & F23, D23, B23, H23\\[1ex]
CEERS-SW-z11-26592 & 14:19:28.72 & $+$52:50:37.15 & $11.4^{+0.4}_{-0.3}$ & $-$20.14 & - & F23\\[1ex]
CEERS-SW-z11-3961 & 14:19:24.03 & $+$52:48:29.00 & $11.2^{+0.3}_{-0.5}$ & $-$19.54 & - & D23, F23, B23\\[1ex]
CEERS-SW-z11-20687 & 14:19:00.64 & $+$52:50:11.98 & $10.9^{+0.3}_{-0.7}$ & $-$19.50 & - & B23\\[1ex]
CEERS-SW-z11-51314 & 14:19:26.62 & $+$52:52:52.18 & $10.8^{+0.5}_{-0.2}$ & $-$19.80 & - & -\\[1ex]
CEERS-SW-z11-1099004 & 14:19:30.20 & $+$52:54:48.81 & $10.7^{+0.3}_{-0.2}$ & $-$20.02 & - & -\\[1ex]
CEERS-SW-z14-37967 & 14:19:27.31 & $+$52:51:29.23 & $14.0^{+0.5}_{-1.7}$ & $-$19.56 & - & D23\\[1ex]
CEERS-2-1-z12-20465 & 14:20:36.36 & $+$52:58:31.70 & $11.5^{+0.2}_{-0.2}$ & $-$20.47 & - & -\\[1ex]
CEERS-2-3-z12-8536 & 14:19:33.58 & $+$52:52:53.37 & $12.1^{+0.3}_{-0.3}$ & $-$20.37 & - & -\\[1ex]
CEERS-2-4-z11-10389 & 14:19:32.62 & $+$52:49:08.57 & $10.8^{+0.5}_{-0.3}$ & $-$19.69 & - & -\\[1ex]
CEERS-2-4-z11-1032400 & 14:19:43.67 & $+$52:50:30.73 & $10.6^{+0.2}_{-0.3}$ & $-$20.41 & - & -\\[1ex]
CEERS-2-6-z11-570 & 14:19:12.14 & $+$52:43:31.87 & $10.5^{+0.5}_{-0.5}$ & $-$19.68 & - & -\\[1ex]
DDT2756-z11-1001979 & 00:14:33.03 & $-$30:26:56.99 & $10.6^{+0.2}_{-0.3}$ & $-$20.53 & 1.20$^{\star}$ & -\\[1ex]
DDT2756-z12-1004647 & 00:14:32.13 & $-$30:26:25.87 & $11.5^{+0.2}_{-0.4}$ & $-$19.67 & 1.35\phantom{o} & -\\[1ex]
DDT2756-z12-1023292 & 00:14:34.98 & $-$30:24:04.00 & $11.7^{+0.5}_{-0.4}$ & $-$18.80 & 1.31\phantom{o} & -\\[1ex]
GLASS-z11-1481 & 00:14:04.53 & $-$30:22:43.10 & $11.1^{+0.4}_{-0.6}$ & $-$18.49 & 1.84\phantom{o} & -\\[1ex]
GLASS-z11-17225 & 00:14:01.76 & $-$30:20:35.53 & $10.7^{+0.3}_{-0.6}$ & $-$19.08 & 1.38\phantom{o} & -\\[1ex]
GLASS-z11-29748 & 00:13:48.34 & $-$30:19:18.50 & $10.7^{+0.2}_{-0.3}$ & $-$20.08 & 1.20$^{\star}$ & C23 \\[1ex]
GLASS-z11-30367 & 00:13:48.33 & $-$30:19:14.62 & $10.6^{+0.2}_{-0.3}$ & $-$19.96 & 1.20$^{\star}$ & C23 \\[1ex]
GLASS-z11-12329 & 00:14:03.30 & $-$30:21:05.64 & $10.6^{+0.4}_{-0.5}$ & $-$19.00 & 1.50\phantom{o} & C23, H23\\[1ex]
GLASS-z14-33570 & 00:13:57.62 & $-$30:18:53.50 & $13.6^{+1.0}_{-1.1}$ & $-$19.40 & 1.20$^{\star}$ & B23\\[1ex]
GLASS-z12-28072 & 00:13:59.76 & $-$30:19:29.15 & $12.2^{+0.2}_{-0.1}$ & $-$20.62 & 1.20$^{\star}$ & C23, N22, D23, H23, B23 \\[1ex]
GLASS-z12-9974 & 00:14:03.26 & $-$30:21:24.54 & $12.1^{+0.6}_{-0.6}$ & $-$19.04 & 1.57\phantom{o} & -\\[1ex]
J1235-z11-2056758 & 12:35:48.19 & $+$04:55:54.84 & $10.9^{+0.3}_{-0.6}$ & $-$19.56 & - & -\\[1ex]
J1235-z12-8807 & 12:35:44.96 & $+$04:54:34.76 & $11.8^{+0.4}_{-0.3}$ & $-$18.94 & - & -\\[1ex]
MACS0647-z11-26400 & 06:47:55.47 & $+$70:15:38.13 & $10.8^{+0.1}_{-0.2}$ & $-$20.37 & 2.53\phantom{o} & C13\\[1ex]
MACS0647-z11-20148 & 06:47:55.74 & $+$70:14:35.81 & $10.6^{+0.2}_{-0.4}$ & $-$20.10 & 5.69\phantom{o} & C13\\[1ex]
NEPTDF-z11-12882 & 17:22:57.72 & $+$65:46:24.54 & $11.3^{+0.3}_{-0.4}$ & $-$19.84 & - & -\\[1ex]
Quintet-z12-9091 & 22:35:51.43 & $+$33:54:32.81 & $12.2^{+0.2}_{-0.3}$ & $-$21.44 & - & -\\[1ex]
SMACS0723-z12-7442 & 07:22:56.36 & $-$73:29:00.52 & $11.7^{+0.3}_{-0.3}$ & $-$19.47 & 1.20$^{\star}$ & D23\\[1ex]
UNCOVER-z11-48395 & 00:14:21.62 & $-$30:21:34.96 & $10.6^{+0.1}_{-0.1}$ & $-$19.70 & 1.89\phantom{o} & -\\[1ex]
UNCOVER-z11-53033 & 00:14:25.36 & $-$30:21:09.45 & $10.8^{+0.4}_{-0.2}$ & $-$19.94 & 1.47\phantom{o} & -\\[1ex]
UNCOVER-z11-2037921 & 00:14:31.07 & $-$30:23:02.76 & $11.0^{+0.2}_{-0.3}$ & $-$19.43 & 1.40\phantom{o} & -\\[1ex]
UNCOVER-z12-30657 & 00:14:09.79 & $-$30:22:49.48 & $11.9^{+0.2}_{-0.2}$ & $-$19.42 & 3.25\phantom{o} & -\\[1ex]
UNCOVER-z12-12816 & 00:14:14.10 & $-$30:24:16.27 & $12.0^{+0.3}_{-0.4}$ & $-$18.83 & 1.73\phantom{o} & -\\[1ex]
UNCOVER-z12-30621 & 00:14:09.80 & $-$30:22:50.08 & $12.3^{+0.2}_{-0.3}$ & $-$19.07 & 3.19\phantom{o} & -\\[1ex]
UNCOVER-z12-6502 & 00:14:23.42 & $-$30:24:54.47 & $11.6^{+0.3}_{-0.3}$ & $-$18.07 & 5.00\phantom{o} & -\\[1ex]
UNCOVER-z12-53293 & 00:14:11.21 & $-$30:21:08.56 & $12.0^{+0.4}_{-0.4}$ & $-$19.06 & 1.81\phantom{o} & -\\[1ex]
UNCOVER-z14-2019859 & 00:14:17.00 & $-$30:24:05.56 & $13.6^{+0.6}_{-1.2}$ & $-$18.76 & 2.51\phantom{o} & -\\[1ex]
WHL0137-22312 & 01:37:18.46 & $-$08:24:26.34 & $11.1^{+0.3}_{-0.5}$ & $-$20.00 & 1.20$^{\star}$ & Br23\\
\hline
\end{tabular}
\caption{Our final sample of $z\geq10.5$ galaxies across all of the fields studied. We include the photometric redshift and intrinsic rest-frame UV magnitude $M_{1500}$, as well as the magnification $\mu$ if applicable. We also include a references column if the candidate has appeared elsewhere (C13=\citealt{Coe2013}, C23=\citealt{Castellano2022,Castellano2023}, Br23=\citealt{Bradley2023}, N22=\citealt{Naidu2022}, B23=\citealt{Bouwens2023}, D23=\citealt{Donnan2023}, F23=\citealt{Finkelstein2023}, H23=\citealt{Harikane2023}).\\$^{\star}$ Magnification is fixed to a floor of 1.2 due to a lack of lensing map coverage.}
\label{tab: candidates2}
\end{table*}

In the Appendix, we provide a field-by-field analysis of the high-redshift candidates that we have selected, where we draw comparisons with other works in the literature. Although we do not search the HUDF for candidates in this particular study, we performed a search for $z>9.5$ galaxies in \citet{Donnan2023b} very recently using the JEMS \citep{Williams2023udf} data. We include the six robust F182M-selected candidates found over one NIRCam pointing in our UV LF determination in Section \ref{Section5}.

\section{UV Luminosity Function}
\label{Section5}
In this section, we present our determination of the UV luminosity function at $9.5\leq z \leq 12.5$. We first describe our methods for assigning survey areas, effective volumes, completeness corrections and uncertainties in galaxy number densities. We then calculate and fit the UV LF, before comparing our new results with relevant determinations from the literature.

\subsection{Survey Area Definition}
As this study is conducted over many inhomogeneous fields, care must be taken in order to define an accurate overall survey area. In particular, the fields with clusters are subject to the effects of gravitational lensing. At the faintest luminosities, the differences between lensing models has been shown to have a potentially large impact on the faint end of the UV luminosity function, with the high magnification regime ($\mu$\,$>$10) more subject to systematic uncertainties (see e.g. \citealt{Bouwens2017}). Moreover, the cluster regions are typically significantly shallower than the survey global depth due to enhanced background from intracluster light and source crowding. For each of our cluster fields, we therefore mask out these shallow cluster regions. Any candidates positioned on these masks are excluded from the UV LF calculation.

\begin{figure*}
\includegraphics[width=2\columnwidth]{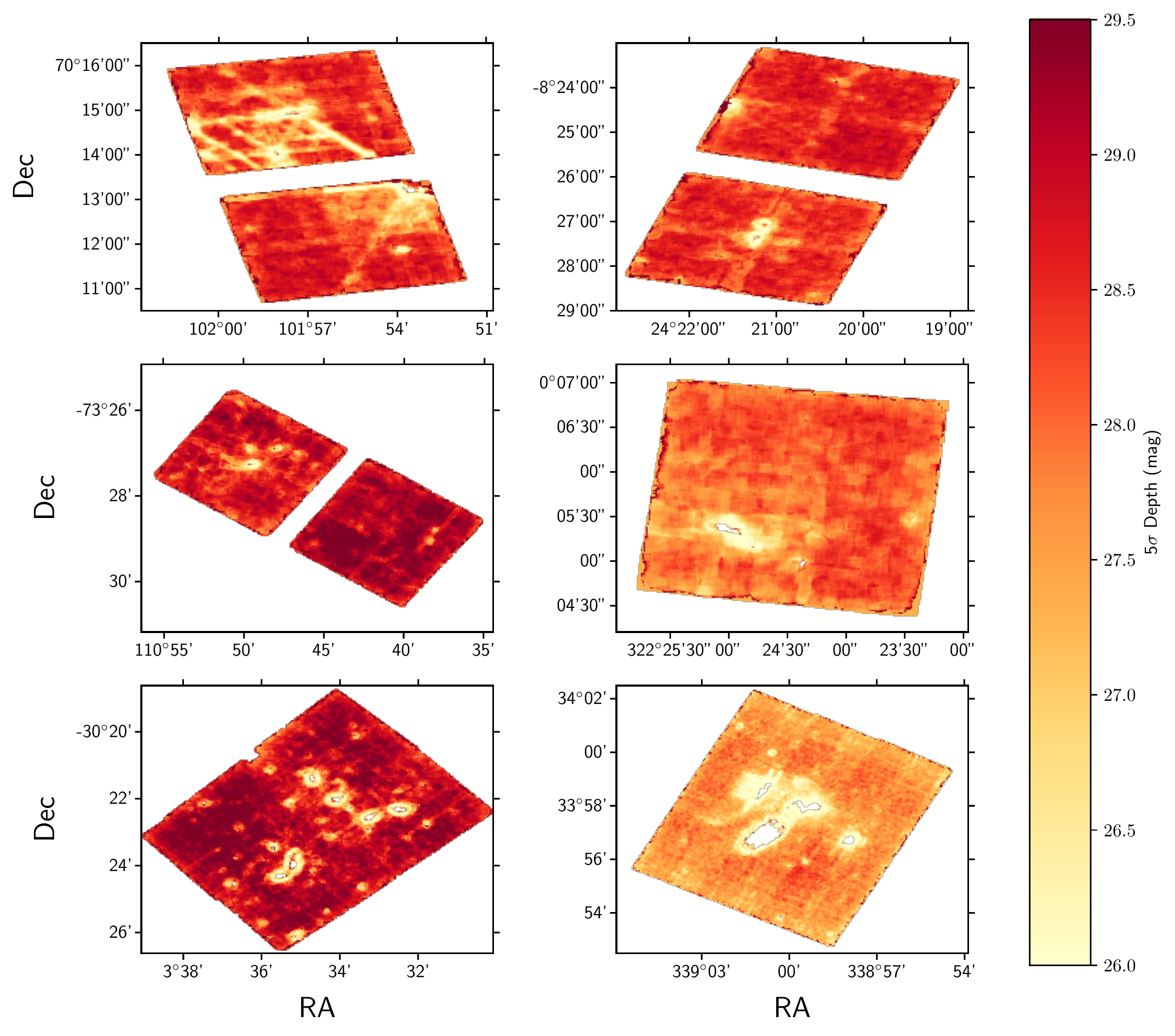}
\caption{Depth maps (measured in 0.35$^{\prime\prime}-$diameter apertures) for the cluster fields and Stephan's Quintet, demonstrating the inhomogeneities in depth across the imaging. The top row is MACS 0647 (left) and WHL 0137 (right), the middle row is SMACS 0723 and RXJ 2129, and the bottom row is UNCOVER and Stephan's Quintet.}
\label{fig:depth_map}
\end{figure*}

For the cluster surveys, we subtract the clusters in a fairly simple way. We first measure the global depth $\sigma_{g}$ across the whole field of view, as well as local depths across the survey pixel by pixel, and produce a depth map as shown in Fig. \ref{fig:depth_map}. We then mask out all pixels that have a local depth that is shallower than $2 \times \sigma_{g}$, removing the cluster and any regions particularly impacted by large foreground galaxies or stars. We then re-calculate $\sigma_{g}$ as our final global depth for the survey. We adopt the same strategy for the fields that are hampered by large foreground objects, e.g. the Cartwheel and Quintet fields.

We de-lens each of the cluster surveys using the lensing models described in Section \ref{surveys_section}. To measure effective survey areas, we calculate the magnification map taking the mid-point of our redshift bin, $z=11$. However, for each candidate, we calculate $\mu$ by generating a magnification map using its $z_{\mathrm{phot}}$. Where there are multiple lensing models considered (i.e. for WHL 0137, MACS 0647 and RXJ 2129), we perform this calculation for each lensing model, and then take the median $\mu$ and de-lensed area. For SMACS 0723 and Abell2744 (UNCOVER, GLASS and DDT 2756) we simply use the magnification and areas calculated using the single lensing model.

For RXJ 2129, we only have one module on the cluster; however for the three surveys with cluster+parallel survey configurations, the lensing maps typically do not to extend to the parallel module, and so we lose the ability to measure the magnification $\mu$. For cluster+parallel pairs, and for the GLASS and DDT-2756 parallels to Abell 2744, we adopt a constant value of $\mu$=1.2 where we do not have any lensing map coverage. This value is motivated by the typical $\mu$ values found around the edge of the lensing maps of the various surveys. We note that adopting an alternative choice of $\mu$=1.1 reduces the overall number densities at the 1-2\% level, as the parallel regions comprise $<$15\% of the overall survey area.

\subsection{Completeness}
An accurate determination of the UV LF requires that the incompleteness of the galaxy sample is accounted for. This becomes of particular importance close to the detection limit of the data. Due to our adoption of a more stringent 8$\sigma$ detection limit throughout, this becomes less of an issue in this survey. The completeness as a function of apparent magnitude in the F200W imaging (F182M for the JEMS data) is determined using the method described in \citet{Donnan2023}. Synthetic galaxies, with a S\'{e}rsic profile of n=1 and half-light radius $0.4$ kpc, i.e. comparable to the typical profile and sizes of F115W-dropouts found in \cite{Morishita2023} at similar $M_{1500}$, were injected into the F200W imaging across each field from $m_{\rm{F200W}}=25-30$ in steps of 0.05 mag. At each step, 1000-2000 sources (depending on the area of the field) are injected and the rate of recovery measured, producing the completeness as a function of the apparent magnitude in F200W. This process is repeated 50 times for each field. We correct the effective volumes of our candidates by their completeness as determined in this procedure.

In a study of the first Hubble Frontier Fields, \citet{Oesch2015} showed the impact of including shear on completeness simulations, finding that large $\mu$ can affect the relative completeness of sources close to the completeness limit. Having removed much of the area around the cluster regions, the effective area of the lensed fields with significant magnification probed in our study presents a modest fraction of the total search area ($<10$\%). This, and the fact that we are conservative in only selecting SNR$>$8 sources, allows us to be confident that the impact of shear on our overall completeness and effective survey volumes is negligible.

\subsection{The $\mathbf{z=11}$ UV luminosity function}
To determine the UV luminosity function for our $9.5\leq z\leq 12.5$ sample we use the $1/{V_{\rm max}}$ estimator \citep{Schmidt1968}. We only include objects for which our F200W detection is greater significance than $8\sigma_{g}$, where $\sigma_{g}$ is the global depth for a given survey. We split our sample into four number density bins between $M_{1500}=-21.80$ and $M_{1500}=-19.55$. At the bright-end, we scale the $M_{1500}=-22.57$ ground-based LF bin from \citet{Donnan2023} to match our redshift range, add in the area we searched, and include it as a fifth bin. In Table \ref{tab:number_densities} we present our number densities $\phi$ in each $M_{1500}$ bin, along with their errors. To calculate the uncertainty in our number densities, we sum in quadrature the uncertainty found through a bootstrap analysis ($\sigma_{b}$) and cosmic variance ($\sigma_{CV}$) as estimated using the \citet{Trenti2008} cosmic variance calculator. One of the key advantages of using many non-contiguous fields for this study is that the relative uncertainties across all the bins are significantly reduced.

\begin{table}
    \centering
    \begin{tabular}{c|c|c}
    \hline
       z & $M_{1500}$ & $\phi$ / $10^{-5}$ Mpc$^{-3}$ mag$^{-1}$\\
        \hline
     \multirow{7}{*}{11.0}  & $-$22.57* & \phantom{*}0.012 $\pm$ \phantom{0}0.012 \\
       & $-$21.80\phantom{*} & \phantom{*}0.129 $\pm$ \phantom{0}0.128 \\
       & $-$20.80\phantom{*} & \phantom{*}1.254 $\pm$ \phantom{0}0.428 \\
       & $-$20.05\phantom{*} & \phantom{*}3.974 $\pm$ \phantom{0}1.340 \\
        & $-$19.55\phantom{*} & \phantom{*}9.863 $\pm$ \phantom{0}4.197 \\
       & $-$18.85* & 23.490 $\pm$ \phantom{0}9.190 \\
       & $-$18.23* & 63.080 $\pm$ 28.650 \\
    \hline
    \multirow{2}{*}{13.5} &   $-$19.45\phantom{*} & \phantom{*}2.469 $\pm$ \phantom{0}1.659 \\
       & $-$18.95\phantom{*} & \phantom{*}6.199 $\pm$ \phantom{0}3.974 \\    
    \hline
    \end{tabular}
    \caption{In the top panel, we present our number density measurements for the $z\simeq11$ UV LF, including those from \citet{Donnan2023} where the absolute magnitude bin has been marked with an asterisk. Note that the $M_{1500}=-22.57$ bin from \citet{Donnan2023} is based on a search over the 1.8 deg$^{2}$ area of COSMOS/UltraVISTA DR5. In the bottom panel, we also include number densities for the $z\simeq13.5$ UV LF.}
    \label{tab:number_densities}
\end{table}

\begin{figure*}
    \centering
    \begin{tabular}{c|c}
    \includegraphics[width=1\columnwidth]{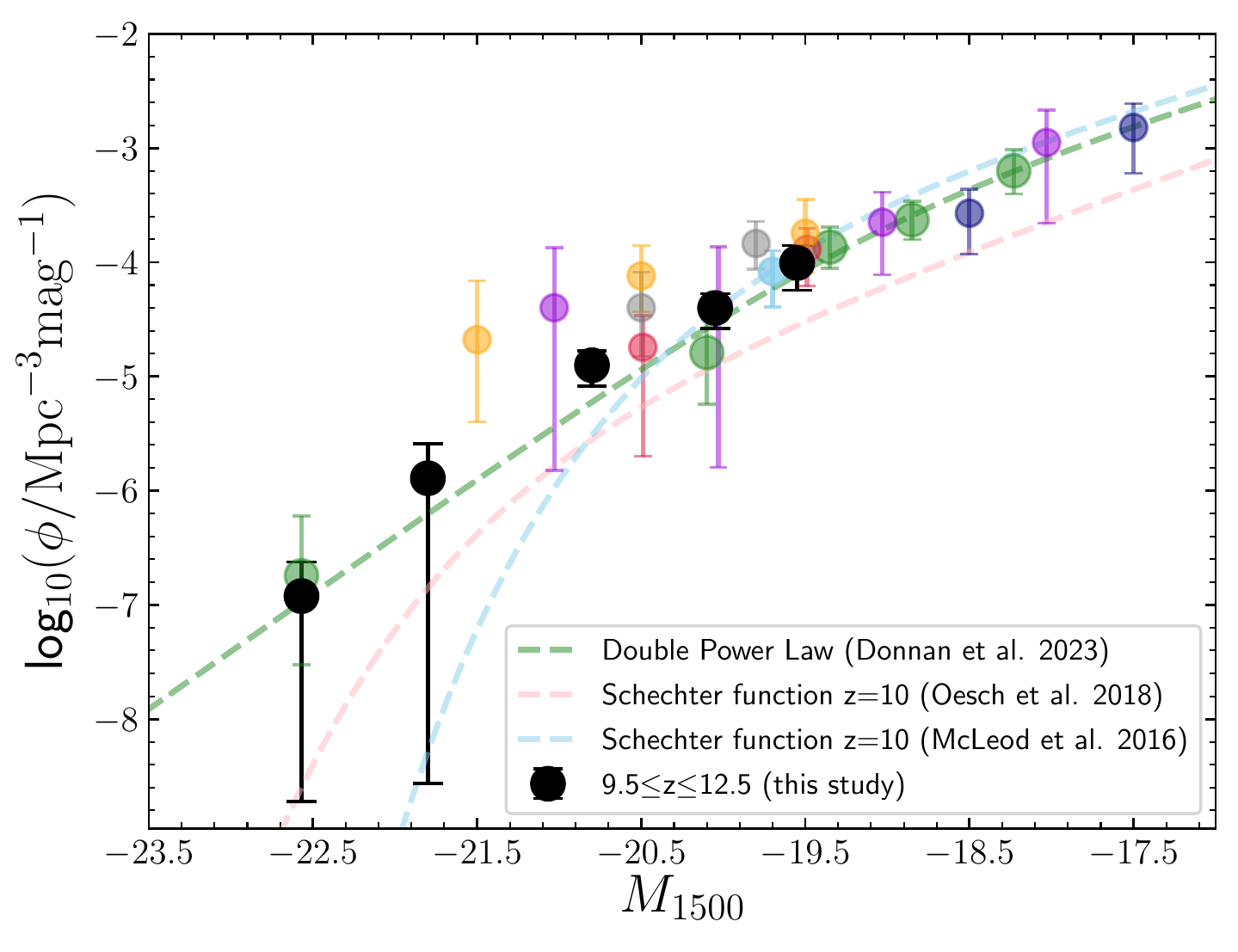} &
    \includegraphics[width=1\columnwidth]{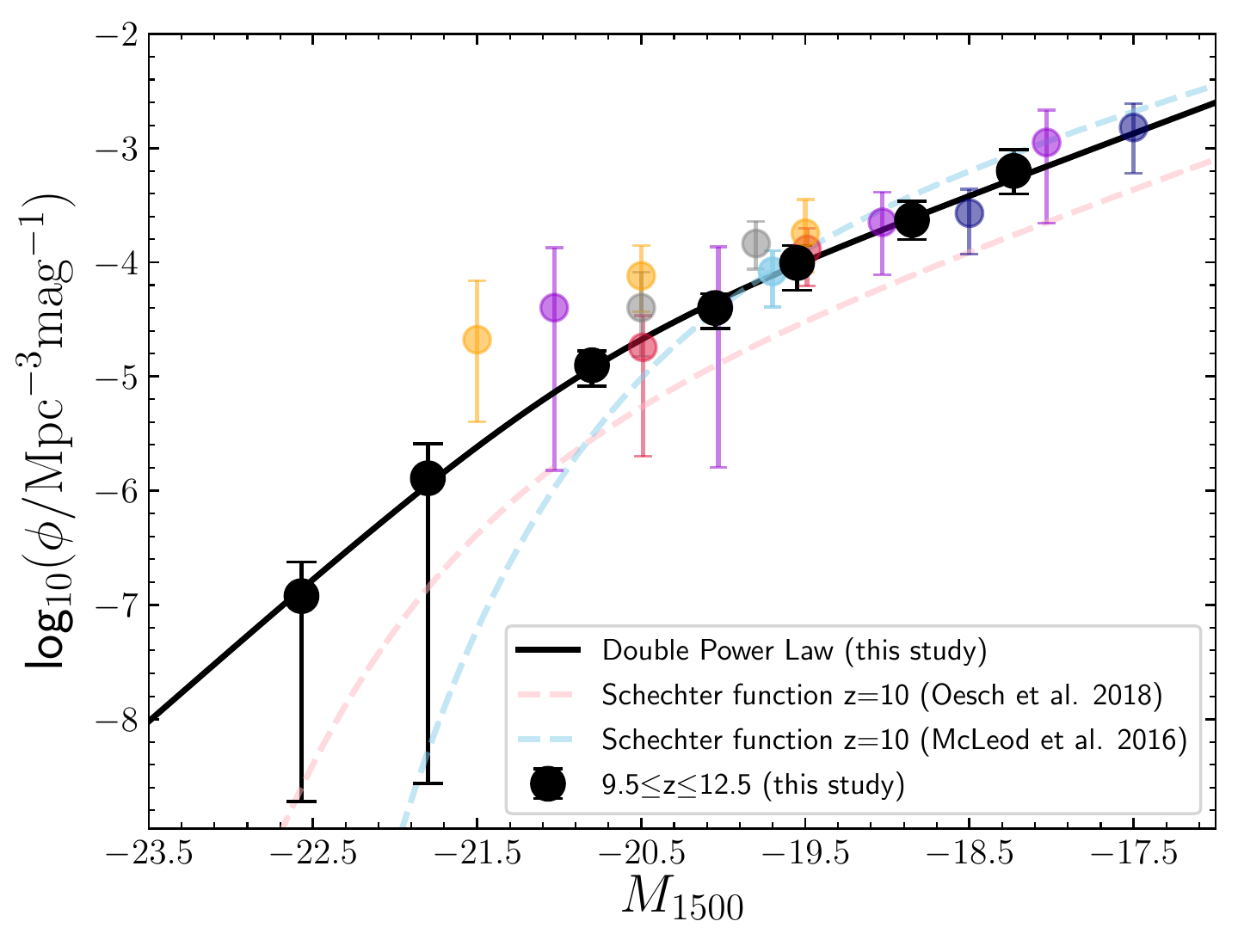}    
    \end{tabular}
    \caption{The UV luminosity function over $9.5<z<12.5$ as found by this study. On the left panel, we include our five number density bins along with data-points from a host of other studies: \citet[][green]{Donnan2023}, \citet[][purple $z=9$]{Harikane2023}, \citet[][orange]{Castellano2023}, \citet[][grey]{Finkelstein2023}, \citet[][red]{Bouwens2023}, \citet[][navy]{Perez-Gonzalez2023}. For the right panel, we have adopted the two fainter $M_{1500}$ bins from \citet{Donnan2023} in order to increase our dynamic range in $M_{1500}$. We have hence fitted a double power law function (DPL) to our LF, adopting as our fiducial DPL fit the fixed $\alpha=-2.35$ case shown in Table \ref{tab: params}. We also plot the \textit{HST}-based Schechter function fits from \citet[][red]{Oesch2018} and \citet[][blue]{McLeod2016}.}
    \label{fig:UVLF11}
\end{figure*}

\begin{figure*}
\centering
\begin{tabular}{c|c}
    \includegraphics[width=1\columnwidth]{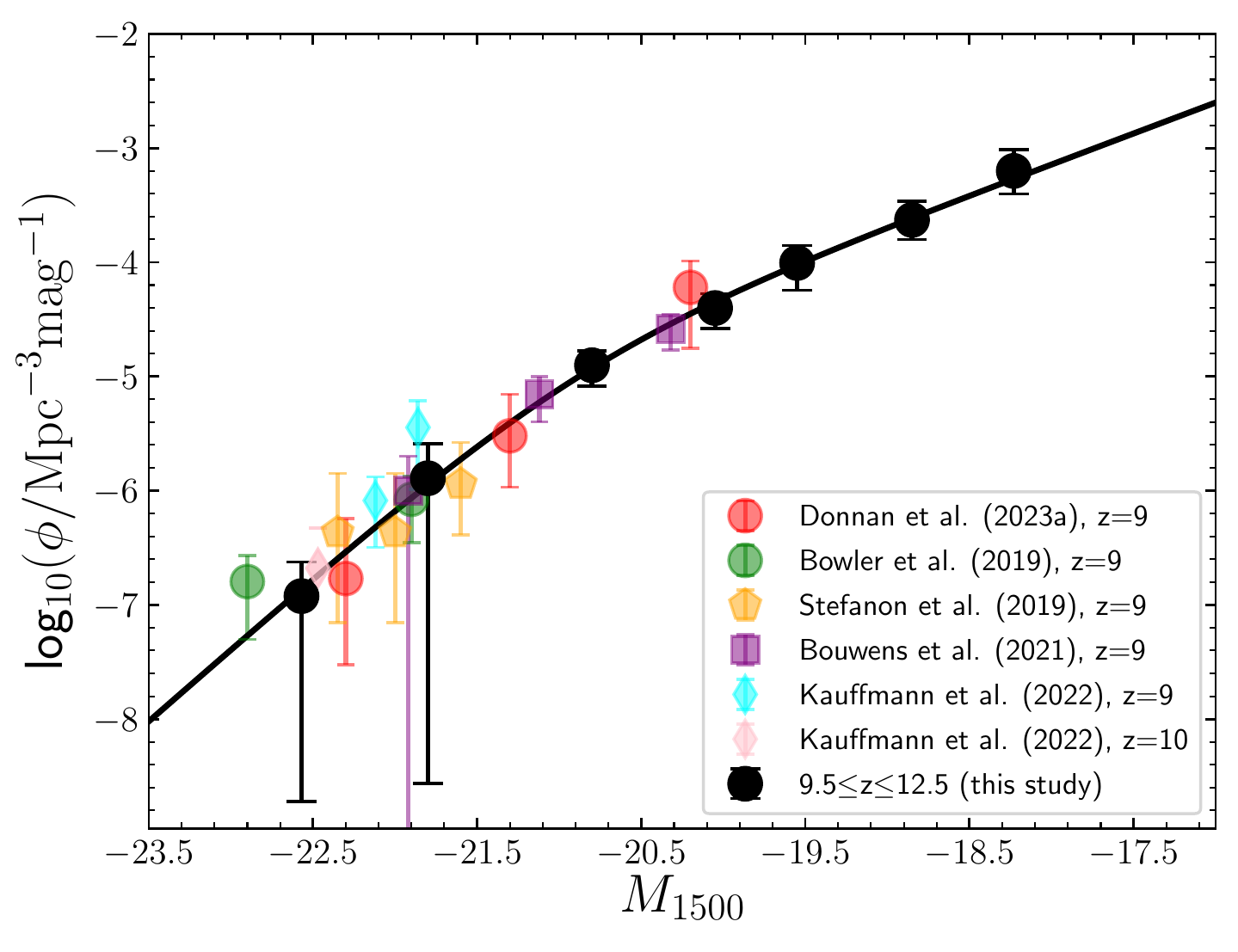} &
    \includegraphics[width=1\columnwidth]{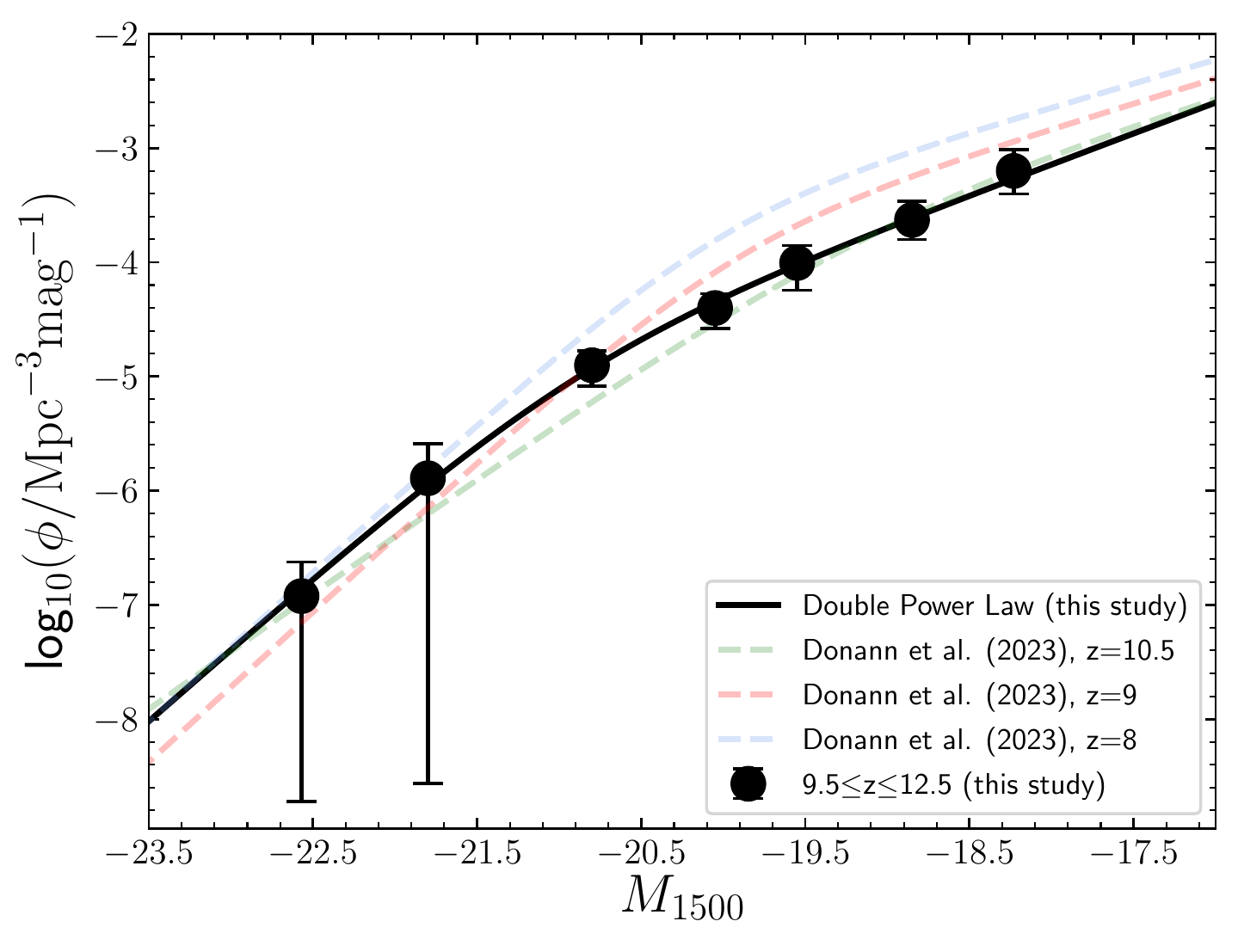}
    \end{tabular}
    \caption{A comparison between our fiducial $z=11$ UV LF and results from other studies. In the left-hand panel we plot a compendium of data points from literature studies at $z=9-10$, whereas the right-hand panel compares our LF with functional fits from \citet{Donnan2023}. Overall, our $z=11$ results appear to be consistent with a lack of evolution in the very bright-end of the LF from $z=9-11$.}
\label{fig:bright-endUVLF11}
\end{figure*}

In the left-hand panel of Fig. \ref{fig:UVLF11} we present our determination of the UV LF at $z=11$, along with determinations across a host of other studies in the literature. There have been numerous LF measurements based on early CEERS epoch 1 and GLASS data, including \citet{Harikane2023}, \citet{Bouwens2023} and \citet{Finkelstein2023}. In the range $M_{1500}=-21.5$ to $-19.5$, we find good agreement with the general consensus in the literature, although we note that they are all over slightly different selection windows: the \citet{Bouwens2023} datapoints are based on $z\simeq9-11$ galaxies, \citet{Donnan2023} is based on a $9.5<z<11.5$ sample and \citet{Finkelstein2023} is over $9.5<z<12.0$. Although we recover all of the candidates in GLASS from \citet{Castellano2023}, our number densities at $M_{1500}\leq -20.5$ are significantly lower. This suggests that the GLASS footprint does cover an overdensity, which is smoothed out by having a much larger overall search area in this study. The faintest $M_{1500}$ bin is in concordance with our earlier work in \citet{Donnan2023} and indeed with earlier \textit{HST} $z=10$ results from \citet{McLeod2016}. However, the excess in the $M_{1500}<=-20.5$ bins presents a clear departure from \textit{HST}-based determinations of the UV LF \citep[e.g.][]{McLeod2016,Oesch2018}.

In order to increase the overall dynamic range, we incorporate the faintest two bins of the UV LF as determined in \citet{Donnan2023}, encouraged by the excellent agreement between our faintest bin and their bin at a similar $M_{1500}$. As well as sharing a similar set of reductions, SED fitting codes and selection criteria, the median redshift of our $9.5\leq z\leq12.5$ sample is $z=10.7$, which is close to the mid-point of their bin at $z=10.5$. The amalgamation of these two studies allows us to determine a UV LF spanning approximately four magnitudes in seven bins, from which we can determine the functional form.

\begin{table*}
\centering
\begin{tabular}{l|c|c|c|c|c|c}
\hline
Function & $M^{\star}$ & $\phi^{\star}$ & $\alpha$ & $\beta$ & $\mathrm \log_{10}{(\rho_{\mathrm{UV}})}$\\
\hline
 DPL free & $-$21.62 $\pm$ 1.28 & (0.43$ \pm $1.15)$\times10^{-5}$ & $-$2.59 $\pm$ 0.35 & $-$4.99 $\pm$ 2.63 & 25.15$^{+0.15}_{-0.18}$ \\[1ex]
 DPL fixed $\alpha$ & $-$20.87 $\pm$ 0.63 & (2.04$ \pm $2.13)$\times10^{-5}$ & $-$2.35 (fixed) & $-$4.16 $\pm$ 0.76 & 25.14$^{+0.11}_{-0.11}$ \\[1ex]
\hline
\end{tabular}
\caption{The parameters determined for our double power law fits to the $z=11$ UV luminosity function. We include both our free fit and the fit fixing $\alpha=-2.35$. We also provide the integrated luminosity density $\mathrm \log_{10}{(\rho_{\mathrm{UV}})}$ for each of the cases, integrated to a limit of $M_{1500}=-17$ (see Section \ref{sfrd_section}).}
\label{tab: params}
\end{table*}

We proceed to fit a double power law, found by \citet{Bowler2014, Bowler2020} and \citet{Donnan2023} to be the preferred functional form at the highest redshifts, rather than a Schechter \citep{Schechter1976} function as found in earlier \textit{HST} studies (e.g. \citealt{McLeod2015,McLeod2016,Bouwens2021}). We fit the seven number density bins both with a free fit and with a fixed faint-end slope. The tabulated double power law fits are presented in Table \ref{tab: params}, and the combined LF and functional forms are shown in the right panel of Fig. \ref{fig:UVLF11}.

We find the free fit double power law is steep at $\alpha=-2.59 \pm 0.35$, however the errors attached to each of the parameters are large, particularly for $M^{\star}$. For the fixed fit, we use $\alpha=-2.35$, which is the $z=10$ slope predicted by the prescription for the evolving double power law as determined by \citet{Bowler2020}. It is also the mid-point between the fixed $\alpha=-2.10$ slope utilised by early \textit{JWST} studies at similar redshifts such as \citet{Donnan2023,Donnan2023b} and \citet{Harikane2023}, and the free-fit value determined in this study. The fit with fixed $\alpha$ mitigates the uncertainties of the free fit significantly, and so we adopt this as our fiducial fit for the functional form of the UV LF at $z=11$ for the rest of this paper. Encouragingly, our fiducial double power law fit appears to be in good agreement with the faintest bin from \citet{Perez-Gonzalez2023}, who probe exceptionally deep imaging of HUDF-par2 as part of their MIRI deep survey of the UDF (PID 1283, PI Oestlin).

It should be noted that alternative fits utilising a range of different fixed faint-end slope are almost indistinguishable over the luminosity range spanned by this study. At $z=8$, \cite{Donnan2023} fitted a faint-end slope $\alpha=-2.04 \pm 0.29$, leveraged by the excellent \textit{HST}-based constraints of the faint-end which reach $M_{1500}=-17$ \citep{McLure2013b}. Although our free fit double power law suggests some steepening between the $z=8$ faint-end slope and that of our $z=11$ LF, fitting our data with fixed $\alpha=-2.04$ also presents an acceptable fit over the magnitude range spanned by this study. This does leave the evolution of the faint-end slope at $z>8$ an open question, despite all of the recent advances brought with this first tranche of \textit{JWST} surveys. With deeper data sets such as NGDEEP (ID 2079, PI Finkelstein; \citealt{Bagley2023b}), studies will be able to constrain $\alpha$ at $z>10$ to the accuracy made possible at $z=8$ with \textit{HST}.

As well as drawing comparisons to recent \textit{JWST}-based determinations of the UV LF, it is instructive to compare to studies probing the bright-end ($M_{1500}<M^{\star}$) of the UV LF at high-redshift. In the left panel of Fig. \ref{fig:bright-endUVLF11} we plot our fiducial LF but also include a compendium of studies probing the bright-end of the LF, including a suite of ground-based studies \citep{Stefanon2019, Bowler2020, Kauffmann2022, Donnan2023} and combined HST Legacy fields from \citet{Bouwens2023}. We also include a comparison of our double power law fit to those determined by \citet{Donnan2023} in the right-hand panel.

It is clear from Fig. \ref{fig:bright-endUVLF11} that our results suggest there is very little evolution in the bright end of the UV LF between $z=9-11$, a feature previously reported by \citet{Bowler2020} between $z=8-10$. In the left-hand panel, it can be seen that the $z=9-10$ literature points at $M_{1500}<-21.50$ scatter symmetrically around our $z=11$ double power law fit. Moreover, in the right-hand panel, it can be seen that at $M_{1500}<-20.50$ there is little to distinguish our $z=11$ fit from the $z=8-10.5$ fits derived by \citet{Donnan2023}.

A relatively un-evolving bright-end has also been reported by \citet{FinkelsteinCANDELS} over the CANDELS fields (see also \citealt{FinkelsteinBagley2022}). Although there have been some surprisingly high number densities of bright $z>10$ galaxies reported in early \textit{JWST} studies \citep{Castellano2022,Castellano2023}, reports of high number densities of bright $z>8$ galaxies had become commonplace even pre-\textit{JWST}, with studies reporting high number densities with wide-area pure parallel \textit{HST} studies \citep{Rojas-Ruiz2020,BagleyHST}, albeit with a greater potential risk of contamination due to limited filter sets.

The interpretation for greater numbers of bright $z>8$ galaxies has been discussed previously in \citet{Bowler2020}, who suggest that the double power law shape reflects a reduction in dust attenuation and/or inefficient mass quenching. Another explanation could lie in the star-formation efficiency, which may be higher at the bright-end to produce the double power law shape at high-redshift (e.g. \citealt{Harikane2022}). After the early \textit{JWST} studies reporting large number densities of bright $z>8$ galaxies, \citet{Mason2023} presented predictions of $z>8$ LFs under the assumption of 100\% star-formation efficiency (SFE). They demonstrated that a 100\% SFE scenario yields LFs several orders of magnitude higher than current results, i.e. observations thus far are fully consistent with current cosmological frameworks. They conclude that our view may be biased in that we are predominantly witnessing young galaxies that are undergoing bursts of star-formation, rather than a representative population. 

There is also the potential for significant field-to-field variation resulting in skewed number densities at the bright-end, where number statistics are generally low. For example, \citep{FinkelsteinCANDELS} found the CANDELS EGS field to be over-dense, with 7/11 of their bright $8.5<z<10.5$ candidates across the five CANDELS fields coming from that field alone. Two of these candidates were spectroscopically confirmed with MOSFIRE observations to lie at $z=8.7$, EGSY8p7 by \citet{Zitrin2015b} and EGS\_z910\_44164 by \citet{Larson2022}. The EGS overdensity at $z=8.7$ could give rise to ionising bubbles \citep{Larson2022}. Further evidence for an ionising bubble was provided by the discovery of an overdensity of $z\simeq9$ candidates found near EGSY8p7 using CEERS \textit{JWST} NIRCam imaging \citep{Whitler2023}. Also recently, the over-density found in GLASS by \citep{Castellano2023} causes their number densities to lie $\sim3\times$ higher than other \textit{JWST} studies.

This study mitigates some of the field-to-field variation with numerous non-contiguous pointings over a relatively large area, while also bridging some of the disconnect between the ground-based and early \textit{JWST} results at $z>9.5$. Yet more progress at the bright-end of the UV LF can be gained through using wide-area surveys such as PRIMER (ID 1837; PI Dunlop) and COSMOS-Web (ID 1727, PI Casey; \citealt{Casey2023}).

\begin{figure}
\centering
\includegraphics[width=1\columnwidth]{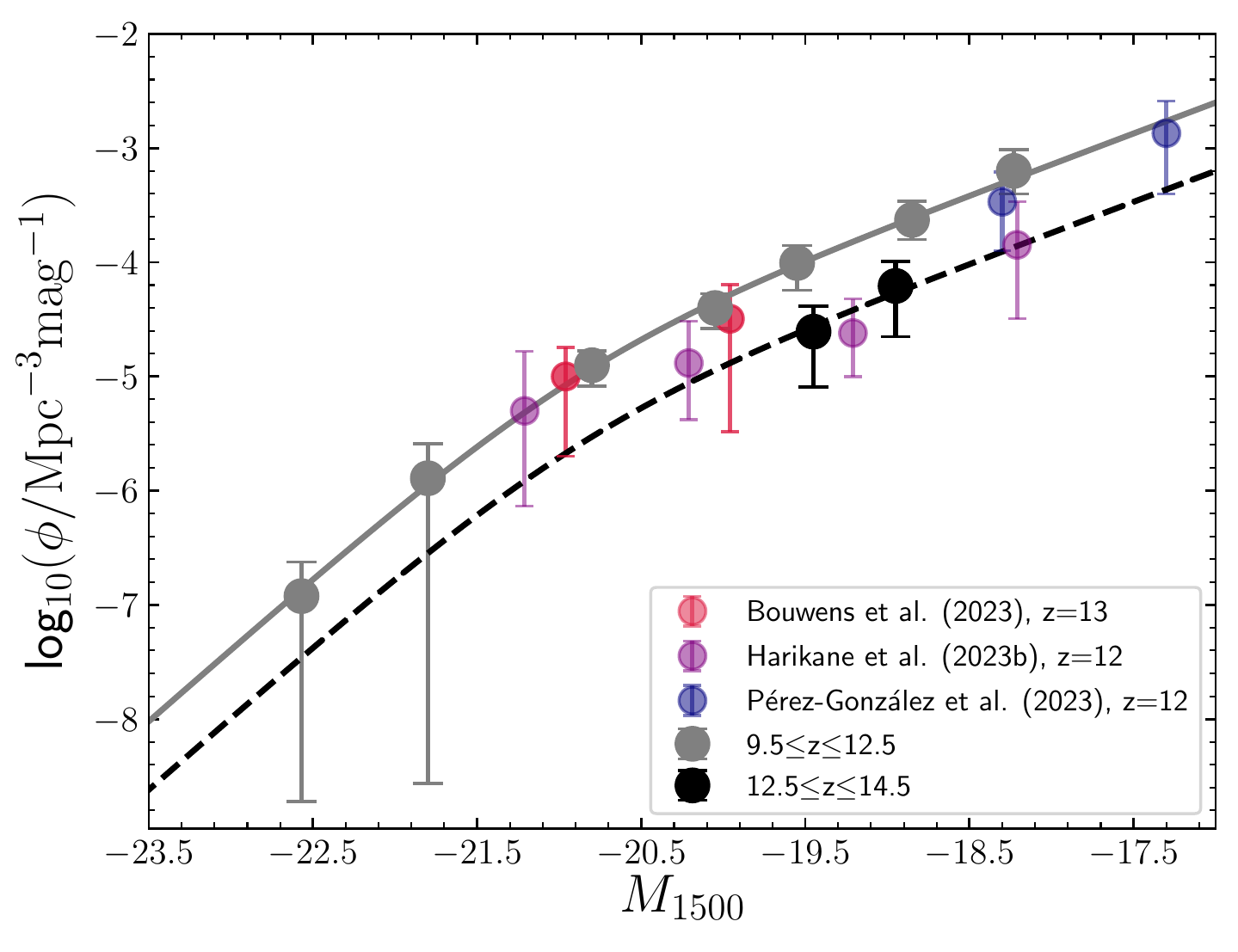}
\caption{Our UV LF determination at $12.5<z<14.5$ (black points). We also include our fiducial $z=11$ UV LF (grey points), and other determinations from the literature at similar redshifts. The dashed black line is our double power law fit to the data, fixing all of the parameters at their $z=11$ values except $\phi^{\star}$. Although the fit is unconstrained at $M_{1500}\leq-20$, precluding any comment on the evolution of the bright-end, we tentatively find that $\phi^{\star}$ falls by a factor of approximately four, i.e. $\phi^{\star} = 5.14 \pm 2.40\times10^{-6}$, with respect to our $z=11$ determination.}
\label{fig:UVLF13}
\end{figure}

Before moving on to the star-formation rate density, we briefly discuss a higher redshift bin generated using our $12.5<z<14.5$ candidates. The number densities are tabulated in Table \ref{tab:number_densities}. As we currently have no constraints on the number densities of $12.5<z<14.5$ galaxies at $M_{1500}\leq-20$, we cannot yet comment on the evolution of the bright-end of the UV LF at these redshifts. However, it is nevertheless instructive to discern how the the number densities evolve between our redshift bins. We hence fit a double power law to our $z\sim13.5$ LF, fixing $\alpha$, $\beta$ and $M^{\star}$ to their fiducial $z=11$ values. This is illustrated in Fig. \ref{fig:UVLF13}. By allowing $\phi^{\star}$ to float, we find a factor of four decrease in $\phi^{\star}$ with respect to our fiducial $z=11$ LF, i.e. $\phi^{\star} = (5.14 \pm 2.40) \times10^{-6}$. Future wide-area searches for bright $z>12.5$ galaxies will be required in order to further determine if the lack of evolution in the bright-end of the UV LF extends to such early times.

\section{Star-formation Rate Density}
\label{sfrd_section}
We proceed to calculate the luminosity-weighted integral for our functional fits to the UV luminosity function in order to determine the UV luminosity density, $\rho_{\mathrm{UV}}$. The integral limit that we choose is $M_{1500}=-17$, as has been convention with previous \textit{HST}-based studies. Our $\rho_{\mathrm{UV}}$ values are tabulated along with the UV LF functions in Table \ref{tab: params}. To convert this to a star-formation rate density, $\rho_{\mathrm{SFR}}$, we use the conversion factor $\cal{K}$$_{\rm UV} = 1.15 \times 10^{-28}$ M$_{\odot}$ yr$^{-1}$/erg s$^{-1}$ Hz$^{-1}$ \citep{Madau2014}. The free-fit and fixed $\alpha=-2.35$ double power law cases yield very similar values of $\rho_{\mathrm{UV}}$. We adopt $\mathrm \log_{10}{(\rho_{\mathrm{UV}})}=25.15^{+0.13}_{-0.14}$ as our fiducial value.

\begin{figure*}
    \centering
    \begin{tabular}{c|c}
    \includegraphics[width=1\columnwidth]{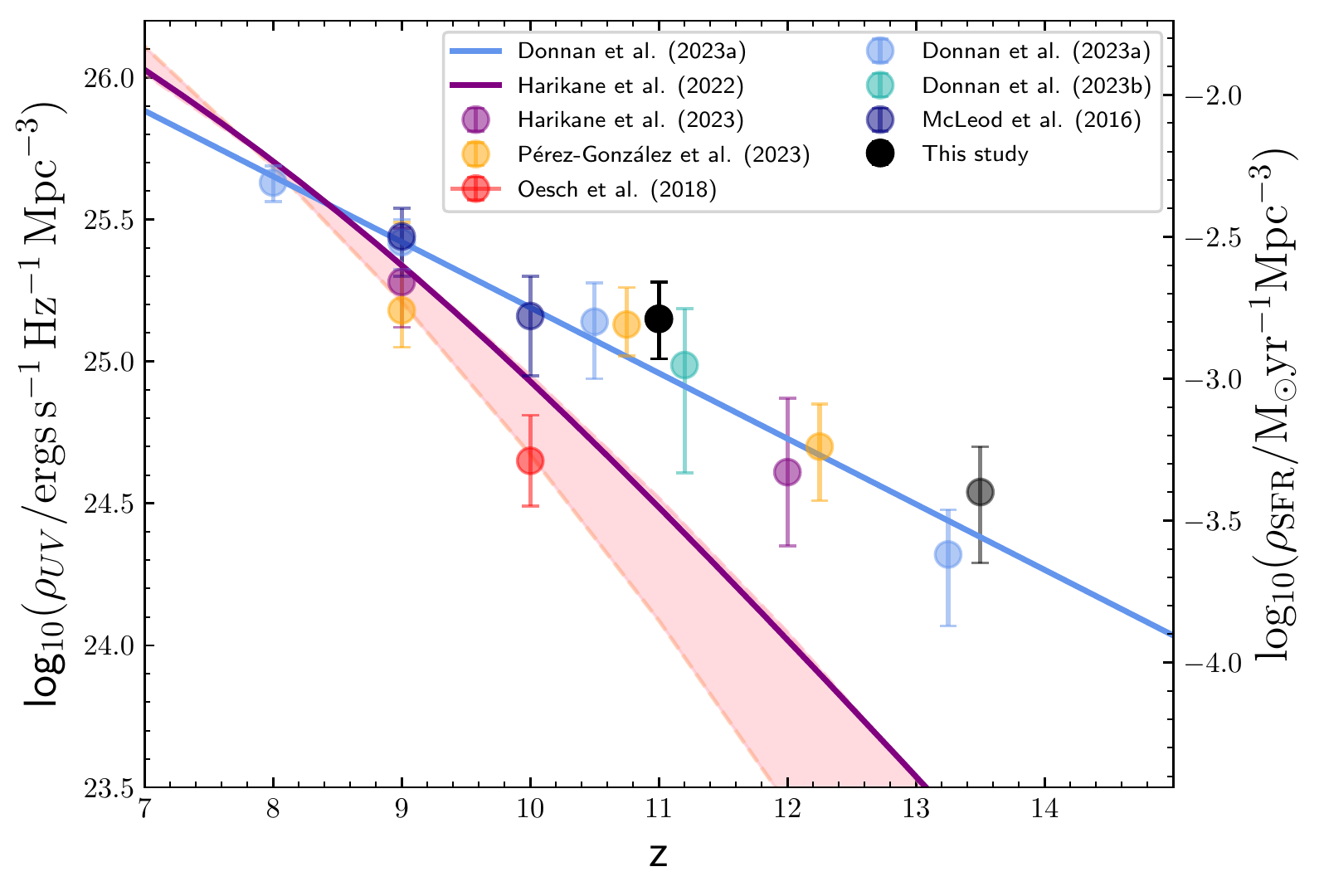} &
    \includegraphics[width=1\columnwidth]{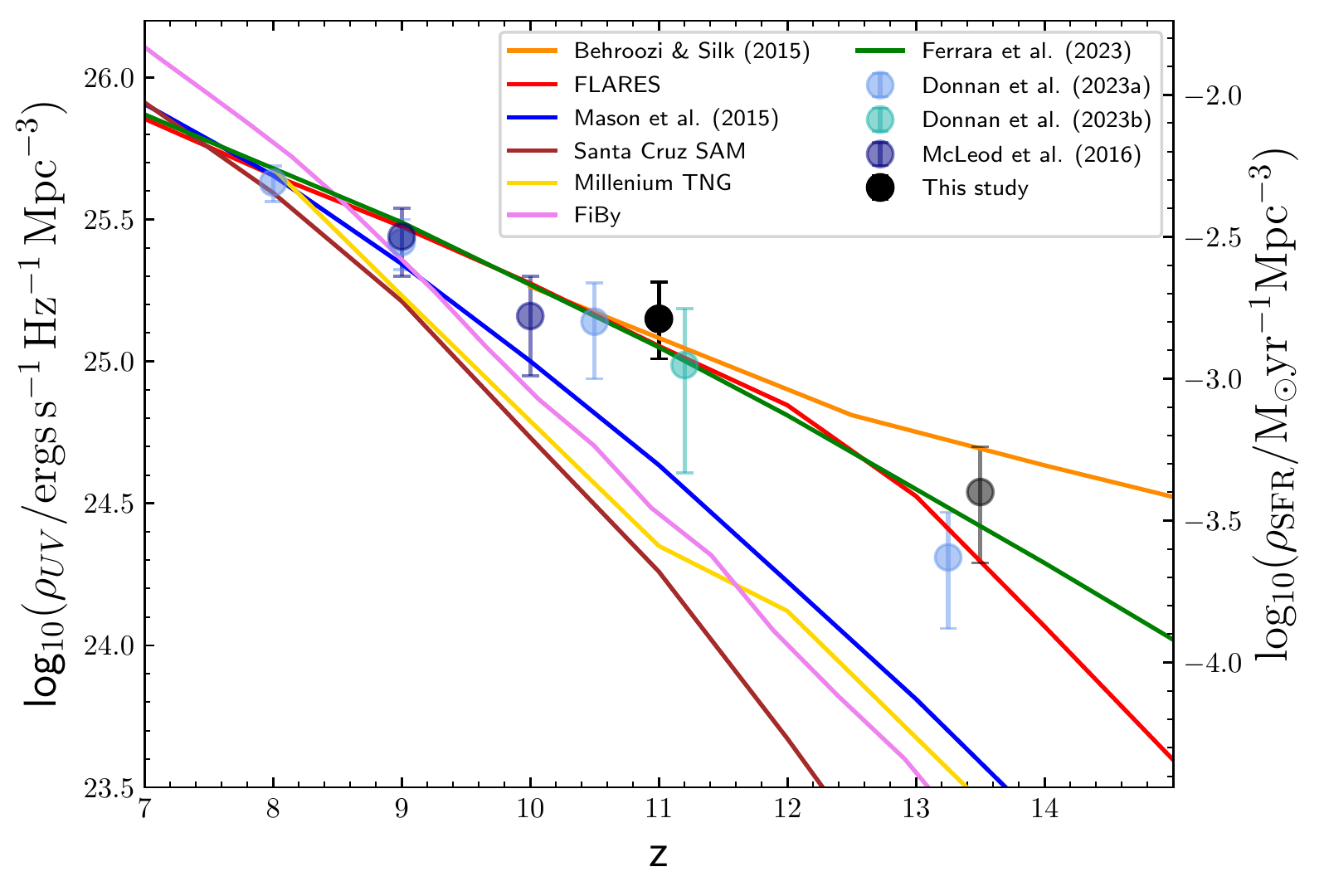} \\
    \end{tabular}
    \caption{Our determination of the luminosity density and star-formation rate density at $z=11$, integrated to a limit of $M_{1500}=-17$. We also include a tentative $z=13.5$ measurement based on our handful of $12.5<z<14.5$ candidates. In the left-hand panel, we include a compilation of early \textit{JWST} studies integrating to the same limit of $M_{1500}=-17$. We also include the \textit{HST}-based determination at $z=10$ from \citet{Oesch2018}, based on the \textit{HST} Legacy Fields, and the rapidly declining $\rho_{\mathrm{SFR}}(z)$ relation based on the evolution of the dark matter halo mass function (pink shading). Finally, we also show the constant star-formation efficiency model (purple line) from \citet{Harikane2022}.
    We find that our new results are slightly higher than, but fully consistent with, the log-linear $\rho_{\mathrm{UV}}(\mathrm z)$ relation (blue line) from \citet{Donnan2023}. In the right-hand panel, we compare our measurements with a suite of theoretical models of galaxy evolution. Typically, the models under-predict the $\rho_{\mathrm{SFR}}$ measurements found both in this study and \citet{Donnan2023,Donnan2023b}, with the exception of the \citet{BehrooziSilk2015}, \citet{Ferrara2023} and FLARES \citet{Wilkins2023} models.}
     \label{fig: SFRD}
\end{figure*}

We also include a tentative measurement of $\rho_{\mathrm{UV}}$ at $z=13.5$, based on our $12.5<z<14.5$ sample. Integrating our $z=13.5$ LF fit yields $\mathrm \log_{10}{(\rho_{\mathrm{UV}})}=24.54^{+0.16}_{-0.25}$ at $z=13.5$.

In the left-hand panel of Fig. \ref{fig: SFRD} we present our determination of $\rho_{\mathrm{UV}}$ along with those from other observations in the literature. Overplotted is the previous \textit{HST}-based results from \citet{Oesch2018} and their rapidly declining $\rho_{\mathrm{SFR}}(\mathrm z)$ relation, following the evolution of the dark matter halo mass function. We also include the constant star-formation efficiency model from \cite{Harikane2022}, which follows $\rho_{\mathrm{SFR}}\propto \mathrm (1+z)^{-0.5}$, and the log-linear relation we previously determined in \citet{Donnan2023}:
\begin{equation}
    \mathrm \log_{10}{(\rho_{\mathrm{UV}})}=(-0.231 \pm 0.037) \,\rm z + (27.5 \pm 0.3)
\end{equation}
As can be seen from the left-hand panel, our new determination of $\rho_{\mathrm{UV}}$, and hence $\rho_{\mathrm{SFR}}$, at $z=11$ is found to be slightly higher than, although still consistent with, our previous estimates from \citet{Donnan2023, Donnan2023b}. With our significantly larger area spread over numerous additional fields, we once again find evidence in support of a smoothly declining $\rho_{\mathrm{SFR}}$ at least to $z=12$. Although tentative, our $z=13.5$ data point also appears to be consistent with the results from \citet{Donnan2023}, \citet{Harikane2023} and \citet{Perez-Gonzalez2023}, and suggests the smooth decline in $\rho_{\mathrm{SFR}}$ may yet continue beyond $z=12$.

Given that the $\rho_{\mathrm{UV}}$ can be sensitive to the assumed faint-end slope, we want to ensure that our result is robust against the assumed $\alpha$ slope, particularly as the current \textit{JWST} data are still unable to robustly constrain this parameter to the same level seen at $z=8$ with \textit{HST}. Hence, we also explore the impact of adopting $\alpha=-2.04$, which was the slope found at $z=8$ by \citet{Donnan2023} and still presents an acceptable fit to our UV LF at $z=11$. We find that $\mathrm \log_{10}{(\rho_{\mathrm{UV}})}(\mathrm z=11) =25.09^{+0.14}_{-0.14}$, which is in good agreement with the log-linear relation from \citet{Donnan2023}. Our conclusion of a smoothly declining $\rho_{\mathrm{SFR}}(\mathrm z)$ between $z=8-11$ is therefore robust against the assumed value of $\alpha$.

Pre-\textit{JWST}, the rate that $\rho_{\mathrm{SFR}}$ declined at $z>8$ was a matter of debate, with \citet{Oesch2018} advocating a rapid decline based on a lack of $z=10$ galaxies found among \textit{HST} Legacy fields. This was a conclusion shared by \citet{Ishigaki2018} when integrating to a limit of $M_{1500}=-17$, the limit of \textit{HST}'s capabilities, although they noted that this depends on the choice of $M_{1500}$ limit, and that a limit of $M_{1500}=-15$ resulted in a smoothly declining $\rho_{\mathrm{SFR}}$. Many theoretical models of galaxy evolution also follow such a trend of rapidly declining $\rho_{\mathrm{SFR}}(\mathrm z)$ beyond $z>8$. In the right-hand panel of Fig.~\ref{fig: SFRD}, we include a comparison between our results and those of theoretical models of galaxy evolution, where $\rho_{\mathrm{UV}}$ has been determined by integrating to the same $M_{1500}$ limit. Our new results lie above a number of theoretical models that follow more closely the rapidly declining relation, such as \citet{Yung2019}, FiBy \citep{Paardekooper2013}, Millenium TNG \citep{Kannan2023} and \citet{Mason2015}. However, the \citet{BehrooziSilk2015}, \cite{Ferrara2023} and FLARES \citep{Wilkins2023} models appear to be entirely consistent with our $z=11$ results.

As well as lying above the \citet{Oesch2018} rapidly declining $\rho_{\mathrm{SFR}}$ relation we also lie above the constant star-formation efficiency model from \citet{Harikane2022}. Although the constant star-formation efficiency model is in close agreement with \citet{Harikane2023} and \citet{Donnan2023} up to $z=9$, the observations begin to diverge from the model at higher redshifts. This is discussed in detail by \citet{Harikane2023}, who show that increasing the star-formation efficiency at $z>10$, or adjusting the initial mass function (IMF) to be more top-heavy, can account for such discrepancies. Alternatively, as was mentioned earlier, \citet{Mason2023} suggest that observations thus far with \textit{JWST} may be biased towards young galaxies with copious star-formation, drawn from the upper envelope of the $M_{1500}-\mathcal{M}_{\mathrm {halo}}$ relation, and that a more representative sample may be found when probing fainter apparent magnitudes, e.g. $m_{200}>30$. 

On the other hand, a study by \citet{Bouwens2023} suggests that perhaps the star-formation rate density may yet lie higher than the results of \citet{Harikane2023}, \citet{Donnan2023} and this study. They compute $\rho_{\mathrm{SFR}}$ when considering a more inclusive (but still referred to as ``solid'') sample of high-redshift candidates, rather than solely the most ``robust'' candidates, and find a significantly higher $\rho_{\mathrm{SFR}}$. Their study highlights the requirement for spectroscopy to confirm or refute the high-redshift nature of candidates. That said, it is worth noting that early \textit{JWST} NIRSpec results from e.g. \citet{Curtis-Lake2023}, \cite{RobertsBorsani2023}, \cite{Bunker2023} and \cite{ArrabalHaro2023} suggest that $z=9-11$ candidate selection with photometric redshifts has been largely successful.

\section{Conclusions}
We have presented a search for high signal-to-noise $z\geq9.5$ galaxy candidates across a suite of twelve \textit{JWST} surveys spanning a raw survey area of $\sim$260 sq. arcmin. We uncover 61 $z\geq9.5$ candidates detected at 8$\,\sigma$, of which eighteen are at $z\geq11.5$. The exceptional brightness of many of these candidates makes them ideal for follow-up spectroscopy with NIRSpec.

With the inclusion of six additional candidates found in our earlier study, \citet{Donnan2023b}, we proceed to construct the UV LF over $9.5<z<12.5$. When combined with faint-end constraints from \citet{Donnan2023}, and UVISTA-based constraints at the bright-end, our $z=11$ LF spans four magnitudes of dynamic range in $M_{1500}$.
Our LF is consistent with previous \textit{JWST}-based number densities in the literature at similar redshifts, but with tighter constraints owing to our wider search area. At the bright-end, our LF is consistent with a lack of evolution between $z=9-11$, similar to results from \citet{Bowler2020}, who found such trends between $z=8-10$. This lack of evolution suggests that we are entering a regime where mass quenching efficiency is lower or that there is reduced dust attenuation \citep{Bowler2020}. 

At the faint-end, our $\alpha$ measurements suggest a steep $\alpha$, which we fix to $\alpha=-2.35$ for our fiducial LF. With the present data, we are not yet able to robustly determine how $\alpha$ evolves between $z=8-11$, given that a fit with $\alpha=-2.04$ (as measured at $z=8$ by \citealt{Donnan2023}) still provides an acceptable fit.

By integrating our LF, we arrive at the luminosity density and hence star-formation rate density. Our $\rho_{\mathrm{UV}}$ at $z=11$ lies slightly above, but is still consistent with, our previous log-linear $\rho_{\mathrm{UV}}(\mathrm z)$ relationship from \citet{Donnan2023}. Our $\rho_{\mathrm{SFR}}$ at $z=11$ suggests a continued smoothly declining $\rho_{\mathrm{SFR}}$ to at least $z=12$, with tentative $z=13.5$ measurements based on a handful of $z>12.5$ candidates further re-inforcing this. This result lies above both the rapidly declining $\rho_{\mathrm{SFR}}$ suggested in pre-\textit{JWST} studies (e.g. \citealt{Oesch2018, Ishigaki2018}) and the constant star-formation efficiency model from \citet{Harikane2022}, as well as numerous theoretical models of galaxy evolution.

There are numerous potential interpretations for the high number densities of bright $z\simeq10$ galaxies and the higher $\rho_{\mathrm{SFR}}$ than predicted by models. At the highest redshifts, we may be seeing a rise in the star-formation efficiency \citep{Harikane2023b}, or that we are potentially viewing the upper envelope of the $M_{1500}-\mathcal{M}_{\mathrm {halo}}$ relation \citep{Mason2023}. Although there have been suggestions that the early \textit{JWST} studies may be subject to significant contamination by low-redshift interlopers, particularly as the overlap in sources between studies has hitherto been relatively low (see discussion by \citealt{Bouwens2023}), we can be encouraged by the number of spectroscopic confirmations of many photometrically selected $z>8$ candidates (e.g. \citealt{Curtis-Lake2023, ArrabalHaro2023, ArrabalHaro2023b}).

The prospects for building upon this LF determination with further \textit{JWST} Cycle 1 surveys are excellent. In the bright-to-intermediate regime, PRIMER will be instrumental in providing homogeneous coverage of COSMOS and UDS at even greater areas than those probed in this study, and to comparable depths. At the very brightest luminosities, COSMOS-Web \citep{Casey2023} will provide $>0.5$ sq. deg. of imaging, which will be crucial for uncovering the brightest (and rarest) high-redshift galaxies. At the faint-end, exceptionally deep surveys such as NGDEEP \citep{Bagley2023b} and JADES \citep{Robertson2023} will help to better constrain the evolution of the faint-end slope $\alpha$. 

\section*{acknowledgements}
We would like thank the referee for very useful comments and a constructive report, which have helped improve the manuscript. We would like to express our appreciation to all of those who have worked tremendously hard to make all of these \textit{JWST} surveys happen. We are also thankful that all of these data sets were made publicly available immediately. 

We would also like to thank all those involved in the creation of the ancillary \textit{HST} data that was publicly available, and to all of those who have provided gravitational lensing models publicly for the community.

We would like to thank Adi Zitrin for feedback on the lensing models utilised, and Marco Lam for useful discussions. We would also like to thank Andrea Ferrara for useful discussion, and for providing additional data from the \cite{Ferrara2023} model.

This work utilizes gravitational lensing models produced by PIs Bradač, Natarajan \& Kneib (CATS), Merten \& Zitrin, Sharon, Williams, Keeton, Bernstein and Diego, and the GLAFIC group. This lens modeling was partially funded by the HST Frontier Fields program conducted by STScI. STScI is operated by the Association of Universities for Research in Astronomy, Inc. under NASA contract NAS 5-26555. The lens models were obtained from the Mikulski Archive for Space Telescopes (MAST).

This research has made use of the SVO Filter Profile Service (http://svo2.cab.inta-csic.es/theory/fps/) supported from the Spanish MINECO through grant AYA2017-84089 \citep{Rodrigo2012} \citep{Rodrigo2020}

This research used the facilities of the Canadian Astronomy Data Centre operated by the National Research Council of Canada with the support of the Canadian Space Agency.

This work is based on observations taken by the RELICS Treasury Program (GO 14096) with the NASA/ESA HST, which is operated by the Association of Universities for Research in Astronomy, Inc., under NASA contract NAS5-26555.

This research has made use of NASA’s Astrophysics Data System Bibliographic Services.

D.\,J. McLeod, C. Donnan, R.\,J. McLure, J.\,S. Dunlop, R. Begley and M.\,L. Hamadouche acknowledge the support of the Science and Technology Facilities Council. A.\,C. Carnall acknowledges the Leverhulme Trust for their support via a Leverhulme Early Career Fellowship. F. Cullen and T. M. Stanton acknowledge support from a UKRI Frontier Research Guarantee Grant [grant reference EP/X021025/1]. R.\,S. Ellis acknowledges funding from the European Research Council (ERC) under the
European Union’s Horizon 2020 research and innovation program (grant agreement No. 669253).

For the purpose of open access, the author has applied a Creative Commons Attribution (CC BY) licence to any Author Accepted Manuscript version arising from this submission.

\section*{data availability}
At the time of writing, the data sets used in this manuscript were all publicly available from the \textit{JWST} archive on MAST and on the CADC.

\bibliographystyle{mnras}
\bibliography{UVLF_z11_McLeod.bib}

\appendix

\section{Discussion of High-redshift candidates}
In the following sections, we discuss the results of our search for high-redshift galaxies field-by-field, along with a (non-exhaustive) comparison to some previous searches in the literature. The variation between \textit{JWST} high-redshift candidate samples has been a theme in the literature (see \citealt{Bouwens2023} for a comprehensive discussion). We stress that differences in the candidate lists can result from a number of factors such as differing reductions, photometry measurement and selection criteria such as non-detection and $\Delta\chi^{2}$ thresholds.

The nomenclature adopted for our final candidate IDs is Field-redshift-catalogueID (e.g. GLASS-z11-1481). When referring to candidates from other studies we use their naming convention.

\subsection{SMACS 0723}
We select one high-redshift candidate from SMACS 0723 in our final sample, SMACS0723-z12-7442.

There have been numerous studies of this cluster field to date, yielding contrasting samples. We only compare candidates that have passed the additional criteria of $F150W-F200W>0.75$ or $F150W<2\sigma$. \citet{Adams2023} reported a sample of four $9<z<12$ galaxies, but only one of which can be compared to our final sample. Their candidate with ID 10234 was initially selected in our sample, where we found $z_{\mathrm{phot}}=11.8^{+0.2}_{-0.1}$. However, there appeared to be a marginal F090W detection upon visual inspection of the imaging, and so we discarded the candidate.

\citet{Atek2023} reported a sample of high-redshift candidates, including two potential $z>15$ candidates and two potential $z=12$ candidates. We do not include their ID z16a due to a $>3\sigma$ detection in F150W, which rules out a potential $z>13$ solution. While we do find a $z_{\mathrm {phot}}=15.6^{+0.5}_{-0.9}$ solution for their ID z16b, we exclude it from our final sample as it has $\Delta\chi^{2}\simeq2$. Both their objects z12a and z12b have $\Delta\chi^{2}\simeq1$ between the high-redshift and low-redshift solutions, so are also excluded.

Most of the SMACS 0723 sample from \citet{Donnan2023} was not included in our final sample due to our more stringent selection and brightness criteria. Their candidate with ID 8347 is in our final sample as SMACS0723-z12-7442. While we find $z_{\mathrm {phot}}=12.2^{+0.5}_{-0.4}$ for their object with ID 1566, it just misses out on being in our final sample through having $\Delta\chi^{2}=3.4$. Finally, their candidate with ID 10566 had been excluded from our final sample due to a $>2\sigma$ detection in the \textit{HST} ACS imaging in our present analysis, potentially due to a noise spike that contaminated the aperture.

\subsection{GLASS}
Using the maximum-depth combination of the epoch1+2 GLASS survey, we report a sample of twelve $z\geq9.3$ galaxy candidates, including five in the range $10.5<z<11.5$ and three at $z>12$. The high density of bright $z\sim10$ galaxies seen in \citet{Castellano2023} is confirmed by this study, as we re-select all five of their candidates at similar redshifts.

With the first epoch of GLASS imaging, \citet{Naidu2022} and \citet{Castellano2022} unveiled two particularly bright $z=10-12$ galaxies. We recover both of these objects in our final sample as GLASS-z10-3283 (GLASS-z10/GHZ1) and GLASS-z12-28072 (GLASS-z12/GHZ2).

A comprehensive study by \citet{Bouwens2023} also covered the GLASS data set. Their final sample includes the two aforementioned candidates in common with our study, \citet{Naidu2022} and \citet{Castellano2022}. We re-select their $z=13.7$ candidate, GLASSP2H-3576218534 in our final sample as GLASS-z14-33570. Although we also recover a high-redshift solution for their candidate ID 4002721259, the $\Delta\chi^{2}$ was insufficient to include in our final sample. Their alternative sample includes an additional $z=13.3$ candidate, GLASSP1H-4015021230, but we cannot confirm its redshift owing to a flat p(z) distribution. Finally, GLASSP1YJ-4016420474 and GLASSP1YJ-4043920397 are $z=9.6$ and $z=9.7$ candidates that we do not include as they appear to have a marginal detection in F115W. This is likely a result of the increased depth when adding epoch 2, as the epoch 1-only version of the data appeared consistent with non-detections for both objects.

\subsection{CEERS}
The first epoch of observations was split into two fields, north-east (NE) and south-west (SW). In CEERS-NE we selected six candidates, among which are CEERS-93316 \citep{Donnan2023} and Maisie's galaxy \citep{Finkelstein2022}. While the former has recently been revealed by spectroscopy to be lower redshift, Maisie's galaxy has been spectroscopically confirmed to be at high-redshift (DDT 2750, PI Arrabal Haro; \citealt{ArrabalHaro2023}). We also note that CEERS-NE-z11-2543 has also been recently spectroscopically confirmed at high-redshift with the same program. Given its new spectroscopic redshift $z_{\mathrm {spec}}=4.9$, we exclude CEERS-93316 from the subsequent analysis. Four of the five remaining candidates were included in \citet{Finkelstein2023}.

There are five galaxies from \citet{Finkelstein2023} that we also find to be $z>9.0$, but do not include in our final sample due to insufficient $\Delta\chi^{2}$ and/or $<8\sigma$ detection. A further six candidates are not included due to $>2\sigma$ detection in F115W, but five of those are consistent with their $z_{\mathrm {phot}}=8-9$ solutions. The only higher redshift candidate excluded by our study for a F115W detection is their candidate ID CEERS-7227, for which we find the F115W to be marginally detected at $2.6\sigma$. We note that \citet{Finkelstein2023} find a $2.2\sigma$ flux for this object in F115W. Only CEERS-NE-z10-3069854 is unique to this study.

In CEERS-SW, we find six more high-redshift candidates, although only two of the final sample are in common with \citet{Finkelstein2023} (our IDs CEERS-SW-z11-26592 and CEERS-SW-z11-3961). Our candidates CEERS-SW-z11-20687 and CEERS-SW-z14-37967 were also found in \citet{Bouwens2023} and \citet{Donnan2023} respectively.

Four objects in the \citet{Finkelstein2023} sample (their IDs 7603, 1748, 2324 and 4012) were also found to have preferred high-redshift solutions in our catalogues, but were detected at $<8\sigma$ so were excluded. Finally, four $z=8-9$ candidates were F115W$>2\sigma$ and so were excluded.

Epoch two of the CEERS observations were split into six NIRCam pointing pairs, and we adopt this naming convention into our candidate IDs. We report twelve high-redshift candidates across the $\sim60$ sq. arcmin. Two of these objects (CEERS-2-5-z10-9617 and CEERS-2-6-z10-3530) have very recently been independently confirmed with spectroscopy to lie at $z\simeq10$ \citep{ArrabalHaro2023b}.

\subsection{Stephan's Quintet}
Our search over the Quintet field yielded one high-redshift candidate over $11.5<z<12.5$. \citet{Harikane2023} also reported two candidates at $z>12$ from the Quintet field. Their $z\sim12$ candidate was in our initial catalogue, and with our photometry appears to be a robust candidate at $z_{\mathrm {phot}}=11.7^{+0.4}_{-0.5}$ and $\Delta\chi^{2}=4.8$. It also passes our additional F150W-F200W colour criteria, but we do not include it in our final sample as it is only detected at $\sim6\sigma$ in F200W and F277W. For the $z=16$ candidate, we obtain a low-z preferred solution of $z=0.6$, with other solutions at $z=16.4$ and $z=4.9$ being of similar probability (within $1\sigma$). We cannot rule out that it is indeed a $z=16.4$ galaxy, but based on the current fits we do not include it in our final sample. Interestingly, the high and low-redshift solutions that we derive are very similar to CEERS-93316. The low-redshift solution with emission lines was ultimately the spectroscopic redshift, suggesting that candidates around $z=16.4$ are vulnerable to this class of contaminant, especially in the absence of photometric constraints from key medium bands such as F250M and F300M.

\subsection{UNCOVER}
From our analysis of the UNCOVER field, we amass a sample of sixteen candidates. We find UHZ1 from \citet{Castellano2023} as UNCOVER-z10-32757 with a similar $z_{\mathrm {phot}}=10.4^{+0.1}_{-0.2}$ and $\Delta\chi^{2}>100$. One additional candidate was a duplicate of GLASS-z10-3283 and so was excluded.

We also find two components from the triply lensed $z\simeq10$ galaxy from \citet{Zitrin2014}, which has now been spectroscopically confirmed to lie at $z=9.8$ \citep{RobertsBorsani2023}. We find JD1B as UNCOVER-z11-15452 which was observed to be exceptionally bright $M_{1500}=-20.9$, but is magnified $\simeq11\times$. JD1A was in our initial catalogue as UNCOVER-z10-15137 with a robust $z_{\mathrm {phot}}=10.3^{+0.3}_{-0.9}$, $\Delta\chi^{2}=17$ solution. This component had a magnification of 13.4 and $M_{1500}=-17.3$. As it was only $7\sigma_{\rm{local}}$ in each of our detection catalogues, it was not propagated to the final sample. The remaining component, JD1C, was of insufficient signal-to-noise to make it into our catalogues at $>5\sigma$, due to enhanced background and crowding.

\subsection{DDT 2756}
In the DDT 2756 parallel field to Abell2744, we found more examples of bright and robust $z\sim10$ candidates, providing more evidence that the region around Abell2744 is over-dense with such galaxies. We report five $9.5<z<11.7$ candidates, including a particularly robust ($\Delta\chi^{2}\simeq43$) candidate with $z_{\mathrm {phot}}=10.6^{+0.1}_{-0.3}$, DDT2756-z11-1001979.

The DDT 2756 survey was also explored in \citet{Castellano2023}, who found one remarkably bright $z\sim10$ candidate, DHZ1. Due to the high-resolution of the F200W imaging, our F200W-detected catalogue had de-blended DHZ1 into two components. The north-east component was selected for our final sample as DDT2756-z10-1010612, and yielded a robust high-redshift solution of $z_{\mathrm {phot}}=9.5^{+0.6}_{-0.3}$. Although the south-west component was excluded from our final sample due to a 2.2$\sigma$ detection in F115W, we performed SED fitting and found an ultra robust $z_{\mathrm {phot}}=9.5\pm0.1$ ($\Delta\chi^{2}>150$) solution, consistent with the north-east component. The system has since been spectroscopically confirmed to be $z=9.31$ \citep{Boyett2023}.

\subsection{WHL 0137}
We report four high-redshift candidates in the WHL 0137 cluster+parallel data set, three of which are $z\simeq9$ candidates and the other with $z=11.1$. There has been one reported search for $z>9$ galaxies over the WHL 0137 data set to date, with \citet{Bradley2023} uncovering a sample of four galaxies over $8.3<z<10.2$. We include in our final catalogue one of the four candidates, WHL0137-z11-22312 (their WHL0137-09319). While we found a similar high-redshift solution for their candidate WHL0137-10060, the $\Delta\chi^{2}$ was insufficient to include this candidate in our final sample. Finally, their WHL0137-08004 and WHL0137-1332 candidates were excluded from our sample due to detections in F115W.

\subsection{MACS 0647}
We report three high-redshift candidates in the MACS 0647 cluster+parallel data set. There has been a detailed analysis of MACS 0647 JD \citep{Coe2013} by \citet{Hsiao2023}. We recover two of the three multiply imaged components of JD: MACS0647-z11-20148 is JD1 and MACS0647-z11-26400 is JD3. Both have extremely robust solutions $\Delta\chi^{2}>100$, and in excellent agreement with the reported value of $z_{\mathrm {phot}}=10.6\pm0.3$ in \citet{Hsiao2023}. This object has since been spectroscopically confirmed to lie at $z=10.17$ by \cite{Harikane2023b}.

We do not include JD2 in our final sample as it has a $2.8\sigma$ detection in F435W, which aligns with \citet{Hsiao2023}'s F435W photometry also being $>2\sigma$. For completeness, we note that we recover a $z=10.8\pm0.2$ solution for JD2 using our photometry, with a $\Delta\chi^{2}=63$ showing it is an extremely robust object despite the F435W SNR$>$2 measurement. For our subsequent LF analysis in Section 5, we only propagate JD1 so as not to double-count multiple images. In addition to MACS0647 JD, we find a notably robust $z=9.5$ galaxy in the cluster module, MACS0647-z9-20158.

\subsection{RXJ 2129}
We report two $z\simeq9$ galaxy candidates in this field, however we do not uncover any $z\geq9.5$ candidates. To date, there has been one reported $z\geq9.5$ galaxy in the RXJ 2129 field with the new \textit{JWST} data \citep{HWilliams2023}. This galaxy is reported to be intrinsically very faint with $M_{1500}=-17.4$, but high magnification factor ($\mu\simeq20$). We do not recover this spectroscopically confirmed galaxy due to its position near the RXJ 2129 cluster. The enhanced background in the region around the object had caused the object to fail our signal-to-noise cuts in all of the catalogues.

\subsection{NEP TDF}
Our study of the NEP TDF data set uncovered one $z\sim10$ candidate and one $z\sim11$ candidate. To date, there have been no published candidates in the literature. \citet{Adams2023b} have recently presented $7.5\leq z \leq 13.5$ luminosity functions based on a search including candidates across this field, but the associated candidate list is yet to be published (Conselice et al. in prep).
\subsection{Cartwheel}
Our final sample does not include any galaxy candidates at $z>9$ from the Cartwheel galaxy field.
\subsection{J1235}
From our study of the J1235 field we report three high-redshift galaxy candidates over $9.7\leq z\leq11.8$. To date, there have been no other published high-redshift candidates in the literature.

\bsp	
\label{lastpage}

\end{document}